\documentclass[fleqn,usenatbib]{mnras}

\usepackage{newtxtext,newtxmath}
\usepackage{multirow}
\usepackage{comment}

\usepackage[T1]{fontenc}
\usepackage{ae,aecompl}

\usepackage{graphicx}    
\usepackage{amsmath}    
\usepackage{amssymb}    

\title[Encounters between a planetary system and binaries]{Encounters involving planetary systems in birth environments: the significant role of binaries}

\author[Li, Mustill \& Davies]{
Daohai Li\thanks{E-mail: li.daohai@astro.lu.se, lidaohai@gmail.com (DL)},
Alexander J. Mustill,
and Melvyn B. Davies
\\
Lund Observatory, Department of Astronomy and Theoretical Physics, Lund University, Box 43, SE-221 00 Lund, Sweden
}

\date{Accepted XXX. Received YYY; in original form ZZZ}

\pubyear{2015}

\hypersetup{draft}
\begin{document}
\label{firstpage}
\pagerange{\pageref{firstpage}--\pageref{lastpage}}
\maketitle

\begin{abstract}
Most stars form in a clustered environment. Both single and binary stars will sometimes encounter planetary systems in such crowded environments. Encounter rates for binaries may be larger than for single stars, even for binary fractions as low as 10-20 per cent. In this work, we investigate scatterings between a Sun-Jupiter pair and both binary and single stars as in young clusters. We first perform a set of simulations of encounters involving wide ranges of binaries and single stars, finding that wider binaries have larger cross sections for the planet's ejection. Secondly, we consider such scatterings in a realistic population, drawing parameters for the binaries and single stars from the observed population. The scattering outcomes are diverse, including ejection, capture/exchange and collision. The binaries are more effective than single stars by a factor of several or more in causing the planet's ejection and collision. Hence, in a cluster, as long as the binary fraction is larger than about 10 per cent, the binaries will dominate the scatterings in terms of these two outcomes. For an open cluster of a stellar density 50 pc$^{-3}$, a lifetime 100 Myr and a binary fraction 0.5, we estimate that of the order of 1 per cent of the Jupiters are ejected, 0.1 per cent collide with a star, 0.1 per cent change ownership and 10 per cent of the Sun-Jupiter pairs acquire a stellar companion during scatterings. These companions are typically 1000s of au distant and in half of the cases (so 5 per cent of all Sun-Jupiter pairs), they can excite the planet's orbit through Kozai--Lidov mechanism before stripped by later encounters. Our result suggests that the Solar System may have once had a companion in its birth cluster.

\end{abstract}

\begin{keywords}
celestial mechanics -- planet-star interactions -- planetary systems -- open clusters and associations: general
\end{keywords}


\defcitealias{Li2019}{Paper I}
\defcitealias{Li2020a}{Paper II}
\section{Introduction}\label{sec-intro}
Most stars are born in clusters together with many other stars. In the solar neighbourhood, the majority of the stars in embedded clusters form in ones of at least 100 members; and, the total stellar mass as a function of the mass of the parent cluster is roughly flat, meaning that similar numbers of stars form in low-mass and high-mass clusters \citep{Lada2003}. Due to gas removal, small clusters quickly become unbound. About 90\% of the stars are formed in clusters that dissolve in $\lesssim10$ Myr \citep{VandenBergh1981,Elmegreen1985,Battinelli1991,Ward2020}. In these short-lived clusters, stars may experience some interactions with other members in the first few Myr \citep{Proszkow2009,Jaehnig2015}.

In this paper, we consider sizeable open clusters of a few 100s to a few 1000s of member stars. Clusters of this size range are long lived with lifetimes of 100s of Myr or more \citep{Lamers2006}. It has been established for such open clusters, an average star encounters another at a distance closer than 1000 au a few times in a few 100s of Myr \citep[e.g.,][]{Malmberg2007}.

Thus, planets orbiting stars in clusters are inevitably subject to the disturbance of these encounter flybys. The close flybys may cause the immediate ejection, capture or orbital excitation of the planetary orbits or in the long-term, induce strong interplanetary interactions in multi-planet systems that lead to loss of planets \citep[for instance][]{Laughlin1998,Malmberg2011,Hao2013,Cai2017,Fujii2019}. In \citet[][hereafter \citetalias{Li2019}]{Li2019} and \citet[][\citetalias{Li2020a}]{Li2020a}, we modelled in detail the encounters between a planetary system and a single star or another planetary system, emphasising the importance of the architecture of the planetary system.

It is well known that in the field, a significant proportion of the stars are not alone in the sense that they form binary, triple or higher multiplicities. For instance, about 50\% of nearby FGK-stars have at least one companion \citep{Duquennoy1991,Raghavan2010}. It may then be tempting to deduce that scatterings between binaries and planetary systems should be not uncommon in clusters. However, the fraction of binaries of the stellar population therein is not necessarily as high because binaries may be destroyed during encounters with other objects \citep{Hut1983}, while some field star binaries may form when pairs of comoving stars leave the cluster \citep{Kouwenhoven2010}.

On the other hand, observations have shown that the binary fraction in the Pleiades cluster \citep{Bouvier1997,Richichi2012} appears to be similar to or higher than the field. This seems to hold for a few other open clusters in general \citep[e.g.,][]{Bica2005,Sollima2010}. Adding to the complexity, the binary population in a cluster is evolving as it ages. For denser globular clusters, there is an anti-correlation between the cluster's age/mass and the binary fraction \citep{Sollima2007,Milone2008}. This trend is less obvious for open clusters; nonetheless, the binary fraction is probably decreasing with time because of the breakup of wide binaries. The initial binary fraction has to be significant otherwise the observed binary fraction cannot be reproduced in the later cluster evolution \citep{Kroupa1995,Kroupa2001a}. For a more comprehensive account on this matter, we refer to \citet{Goodwin2007}.

The facts accumulated above suggest that in addition to scatterings between planetary systems and single stars \citep[e.g.,][]{Hills1989,Pfalzner2005,Jilkova2016}, ones involving planetary systems and binaries should be as common. The cross section for the orbital excitation/ejection/capture of the planet has been studied by a series of works and scaling laws derived \citep{Laughlin1998,Adams2001,Adams2006,Li2015}. Recently, \citet{Wang2020} surveyed the binary parameters and considered planets initially in the binary and cross sections estimated, highlighting the effectiveness of binaries in causing the planet's ejection and collision.

Here in this work, we delve into the scattering between a planetary system, as exemplified by the Sun-Jupiter pair, and a binary. We calculate the cross sections for the planet's ejection/capture/collision as well as that for the exchange of the planetary system as a whole into a binary. We pay special attention to the orbital architecture with planets ending up in binaries.

The paper is organised as follows. In Section \ref{sec-grid}, we examine the scatterings between the planetary system and a binary where the binary parameters are on a gird to study the parameter dependence. Then in Section \ref{sec-mont}, the binary properties are drawn from the observed population, allowing us to link the then-calculated cross sections to a realistic cluster environment. Section \ref{sec-solar} is devoted to the implications of the above results for the solar system in its birth cluster. We summarise the main results in Section \ref{sec-con}.

\section{Parametric study}\label{sec-grid}
We first explore the effect of the binary properties on the scattering outcomes where the binaries are generated with parameters on a grid. This set of simulations also serves as a guide for later Monte Carlo simulations where the binaries are created with parameters consistent with the observation (Section \ref{sec-mont}).

The parameter space for a four-body scattering, i.e., that between a star--planet pair and a binary star, is enormous. Here, we fix the former to be the Sun--Jupiter pair -- the Sun orbited by Jupiter at 5 au with zero eccentricity. The binary parameters are picked from a grid as detailed below. In the remaining of the paper, we call the central star S, the planet J (and the two together as the SJ-pair), the binary ``bin'' and the components b1 and b2.

\subsection{Simulation strategy and parameter choice}
All the simulations performed in this work have made use of the publicly available $N$-body package {\small FEWBODY} \citep{Fregeau2004}, a code designed to run scattering experiments between a small number of objects. It integrates the Newtonian gravitational system using a high order Runge-Kutta Prince-Dormand method with variable step sizes. In the simulation, when two objects physically touch each other, the two are merged conserving linear momentum. The simulation is started when the encountering objects are sufficiently far away so that the relative tidal perturbation is small and stopped using a similar threshold. The code also automatically classifies the result, looking for stable binary/triple configurations. Higher hierarchies are detected recursively. In $<$1\% of the cases, the final outcomes cannot be resolved within a hard CPU time limit. These are abandoned in our analysis. Our single step error tolerance is et to $10^{-9}$; the typical error of a run is of the order of $10^{-7}$ and the resulting planetary semimajor, if not perturbed, is conserved well (cf. Figure \ref{fig-ater-ajup}).

This work aims to calculate the cross sections for the different events out of the scatterings. The dynamics of scatterings between a stellar binary and a star-planet pair is rich, giving rise to numerous possible outcomes. The total number of types of the outcome as detected by {\small FEWBODY} is over 80 (see Section \ref{sec-mont}). It is thus infeasible to present each and every of them. We here in this section only discuss two cases: (1) the ejection of Jupiter, meaning that Jupiter is a free-floating planet without a host star and (2) in general when the status of the SJ-pair changes (to be detailed below and cf. Figure \ref{fig-illu}). Further categorisation schemes are deferred to Section \ref{sec-mont}.

Suppose we have carried out a suite of $N$ scattering runs and that the upper limit for the impact parameter $b_\mathrm{max}$ is large enough (in the sense that when $b>b_\mathrm{max}$, no event of interest can happen) and we observe that event X occurs $N_\mathrm{X}$ times. The cross section for X happening is then \citep{Hut1983}
\begin{equation}
\label{eq-cross-def}
\sigma_\mathrm{X}=\pi b^2_\mathrm{max} {N_\mathrm{X}\over N}.
\end{equation}

How do we know whether $b_\mathrm{max}$ is large enough? In the case of encounters between a single star and a binary, usually $b_\mathrm{max}$ is expressed in the unit of binary semimajor axis $a_\mathrm{bin}$ in the form $b_\mathrm{max}=(C v_\mathrm{crit}/v_\mathrm{inf}+D)a_\mathrm{bin}$ \citep{Hut1983}. Here $v_\mathrm{crit}$ is the critical encounter velocity at which the total energy of the system (kinetic plus potential) is zero, $v_\mathrm{inf}$ the encounter velocity, i.e., the relative velocity between the encountering objects when their distance is infinity; $C$ and $D$ are empirically determined constants \citep{Hut1983,Bacon1996}. The critical velocity for our encounter between the SJ-pair and the binary star is \citep[e.g.,][]{Antognini2016}
\begin{equation}
\label{eq-vcrit}
v^2_\mathrm{crit}={G( m_\mathrm{SJ} +m_\mathrm{bin} )\over m_\mathrm{SJ}m_\mathrm{bin}}\left({m_\mathrm{S}m_\mathrm{J}\over a_\mathrm{SJ}}+{m_\mathrm{b1}m_\mathrm{b2}\over a_\mathrm{bin}}\right),
\end{equation}
where $G$ is the gravitational constant; $m_\mathrm{S}$ and $m_\mathrm{J}$ are the masses of the Sun and Jupiter, $m_\mathrm{SJ}$ the sum, and $a_\mathrm{SJ}$ the Jovian semimajor axis; $m_\mathrm{b1}$ and $m_\mathrm{b2}$ are the masses of the two binary components, $m_\mathrm{bin}$ the total, and $a_\mathrm{bin}$ the binary semimajor axis. Obviously, because $m_\mathrm{J}$ is much smaller than the other objects, the Sun-Jupiter binding energy does not contribute much. Therefore, $v_\mathrm{crit}$ is governed by the binary properties. In this case, $v_\mathrm{crit}$ is not a good threshold because it does not encompass much information for the SJ-pair. Perhaps the Keplerian orbital velocity is in some cases better used to scale the encounter velocity \citep{Hills1989,Fregeau2006}. We thus opt to not use the formula by \citet{Hut1983}.

Previous works on the encounters between a star--planet pair and a binary pair sometimes used a fixed ratio between $b_\mathrm{max}$ and $a_\mathrm{bin}$. For example, values such as $b_\mathrm{max}= 2 a_\mathrm{bin}$ and $b_\mathrm{max}= 10 a_\mathrm{bin}$ were adopted \citep[][]{Laughlin1998,Li2015}. However, for very close binaries $a_\mathrm{bin}\ll a_\mathrm{SJ}$, this means that only encounters with very small $b\ll a_\mathrm{SJ}$ were accounted for. But \citet{Hills1989} showed that actually $b_\mathrm{max}\gg a_\mathrm{SJ}$ especially when $v_\mathrm{inf}$ was small since the encounter was highly gravitationally focused.

It seems that for our purpose to explore a diverse range of binary parameters, we do not have a simple ready-to-use recipe for choosing $b_\mathrm{max}$. Therefore, we resort to a recursive procedure, i.e., to progressively increase $b_\mathrm{max}$ until it is large enough.

Suppose we have finished the $j$th iteration and are to perform the ($j+1$)th. In all iterations already done, we have been recording the parameter $b_\mathrm{occ}$, the largest impact factor, at which event X is {\it observed} and also $n_{\mathrm{done},j}$ until the $j$th iteration, the total number of runs (a run is a scattering experiment; whether X happens or not) carried out thus far. In the $j$th iteration, the upper limit $b_{\mathrm{max},j}$ has been used for creating the scatterings and $n_j$ is the number of runs performed in that iteration. Now we need to determine $b_{\mathrm{max},j+1}$ and $n_{j+1}$ for the ($j+1$)th iteration.

We compare $b_\mathrm{occ}$ and $b_{\mathrm{max},j}$. (1) If $b_\mathrm{occ}>0.9 b_{\mathrm{max},j}$, we deem that we may have missed encounters with $b>b_{\mathrm{max},j}$ that still allow X to happen so $b_\mathrm{max}$ needs to be increased. In iteration $j+1$, we let $b_{\mathrm{max},j+1}=1.2 b_{\mathrm{max},j}$. Also, to save CPU time, we now do not sample $b$ in the entire range $(0,b_{\mathrm{max},j+1})$ but instead we only generate encounters with $b\in(b_{\mathrm{max},j},b_{\mathrm{max},j+1})$. Then the expected number of encounters within this range needs to be estimated. In general, the encounters should be fully geometric and the cumulative distribution function (CDF) of $b$ should follow $\mathrm{CDF}\propto b^2$. Hence, for the same CDF until iteration $j$ to hold in ($j+1$) and given that there have been $n_\mathrm{done,j}$ scatterings with $b\in(0,b_{\mathrm{max},j})$, the number of encounters with $b\in(b_{\mathrm{max},j},b_{\mathrm{max},j+1})$ should be $n_{j+1}=(b^2_\mathrm{max,j+1}/b^2_{\mathrm{max},j}-1)n_{\mathrm{done},j}=0.44n_{\mathrm{done},j}$. (2) If $b_\mathrm{occ}<0.9 b_{\mathrm{max},j}$, we think $b_{\mathrm{max},j}$ is large enough and in iteration $j+1$, we let $b_{\mathrm{max},j+1}=b_{\mathrm{max},j}$; encounters are created with $b\in(0,b_{\mathrm{max},j+1})$; the number of runs in this iteration is $n_{j+1}=n_j$.

Now we need to choose $b_\mathrm{max,1}$, the initial guess. This is done by first specifying a closest encounter distance $r_\mathrm{enc,1}$ outside which event X we think can rarely happen. We formulate $r_\mathrm{enc,1}$ as
\begin{equation}
r_\mathrm{enc,1}=2a_\mathrm{SJ} \sqrt[3]{m_\mathrm{bin}\over m_\mathrm{S}+m_\mathrm{J}}+2a_\mathrm{bin} \sqrt[3]{m_\mathrm{S}+m_\mathrm{J}\over m_\mathrm{bin}}.
\end{equation}
Then we use the gravitational focusing to calculate the $b_\mathrm{max,1}$ corresponding to $r_\mathrm{enc,1}$
\begin{equation}
\label{eq-bmax-1}
b_\mathrm{max,1}=r_\mathrm{enc,1}\sqrt{1+{2G( m_\mathrm{SJ} +m_\mathrm{bin} )\over r_\mathrm{enc,1}v^2_\mathrm{inf}}}.
\end{equation}

For the first iteration, $n_{1}=1000$. Then all parameters are pinned down. The iterations will then carry on and are stopped if (1) $n_\mathrm{done}\ge$ 20000 and $b_\mathrm{occ}<0.9 b_\mathrm{max,j}$ where convergence is achieved or (2) $n_\mathrm{done}$ reaches 40000 and $b_\mathrm{occ}>0.9 b_\mathrm{max,j}$ where we assume to obtain a converging result is beyond our computational resources.

The above choice for $r_\mathrm{enc,1}$ and $b_\mathrm{max,1}$ may be somewhat arbitrary. Suppose a situation where our estimate of $b_\mathrm{max,1}$ is too small such that in every iteration, we have $b_{\mathrm{max},j+1}=1.2 b_{\mathrm{max},j}$. Then we always have $n_{j+1}=0.44 n_{\mathrm{done},j}$ and $n_\mathrm{done,j+1}=1.44n_{\mathrm{done},j}$. Hence, $n_{\mathrm{done},j}=1.44^{j-1} n_\mathrm{done,1}$. We note at iteration 1, $n_\mathrm{done,1}=1000$. So the maximum allowed number of iterations is $\mathrm{floor}[\log_{1.44}(40000/1000)]=10$. Correspondingly, $b_\mathrm{max,10}/b_\mathrm{max,1}=1.2^{10}=6$. Thus as long as our initial $b_\mathrm{max,1}$ is not too small, we should obtain convergence and this is the case for all simulations.

Now, we are left with choosing parameters for the binaries. Here we vary the binary total mass $m_\mathrm{bin}$, the mass ratio $q_\mathrm{bin}$, the binary semimajor axis $a_\mathrm{bin}$ and $v_\mathrm{inf}$ as listed in Table \ref{tab-param}. For $v_\mathrm{inf}$, we limit it to $<$2 km/s as we are interested in encounters in open clusters where velocity dispersion is small \citep{Binney2008}. We have also fixed the binary eccentricity $e_\mathrm{bin}$ to be zero for most of our simulations and only in a few cases do we also test $e_\mathrm{bin}=0.5$ and $0.99$.

{\small FEWBODY} also needs a finite radius for each object for the detection of physical collisions. For the Sun and Jupiter, their actual radii are used. For a component of the binary, a mass-radius relation of the form of a broken power $R(M)\propto M^\alpha $ is used, where $\alpha=0.6$ when $M>1$ and 0.8 otherwise \footnote{\url{http://personal.psu.edu/rbc3/A534/lec18.pdf}}.

\begin{table}
\centering
\caption{Binary properties adopted in the gird simulations. The quantity is listed in the first column and the values in the second. For the binary eccentricity, $e_\mathrm{bin}=0$ has been tested for all binary configuration whereas 0.5 and 0.99 are only examined for $q_\mathrm{bin}=1$ and $a_\mathrm{bin}\ge 5$ au. We have also performed a further set of simulation for scatterings between the SJ-pair and a single star, marked by $a_\mathrm{bin}=0$.}
\label{tab-param}
\begin{tabular}{c c}
\hline
binary property&values\\
\hline
total mass $m_\mathrm{bin}$ (solar mass)&0.2, 1, 5, 25\\
mass ratio $q_\mathrm{bin}$ &1, 1/5, 1/25\\
semimajor axis $a_\mathrm{bin}$ (au)&(0) 0.2, 5, 125, 3125\\
eccentricity $e_\mathrm{bin}$ &0 (0.5, 0.99)\\
encounter velocity $v_\mathrm{inf}$ (km/s)&0.5,1,2\\
\hline
\end{tabular}
\end{table}

We have also run a suite of scatterings between the SJ-pair and a single star. The stellar mass and encounter velocity are the same as for the binaries. Singles are marked as $a_\mathrm{bin}=0$ in Table \ref{tab-param}.

\subsection{Maximum $b$ for Jupiter's ejection and the change of the status of the SJ-pair}\label{sec-bocc}

\begin{figure}
\includegraphics[width=\columnwidth]{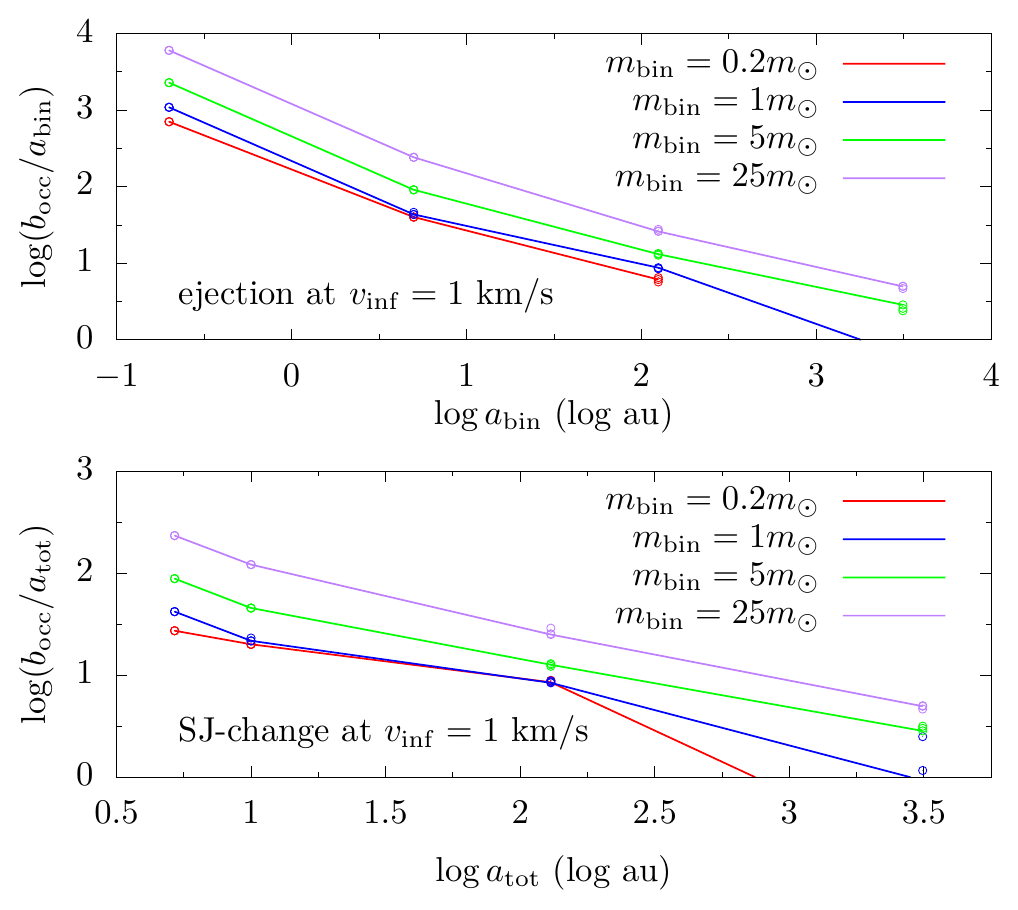}
\caption{The largest impact factor $b_\mathrm{occ}$ observed in the simulations leading to the ejection of Jupiter in the top panel and that of SJ-change (including Jupiter's ejection, collision, capture by other stars, and the exchange of the SJ-pair into a binary) in the bottom. The $x$-axis is logarithm of the binary separation $a_\mathrm{bin}$ in the top panel and the sum of the orbital separations of the SJ-pair and of the binary $ a_\mathrm{tot}=a_\mathrm{SJ}+a_\mathrm{bin}$ in the bottom. Shown here are only runs with $v_\mathrm{inf}=1$ km/s and $b_\mathrm{occ}$ has been normalised against the corresponding $x$ variable. Colours are used to represent different binary masses: $m_\mathrm{bin}=0.2 m_\odot$ in red, $1 m_\odot$ in blue, $5 m_\odot$ in green, and $25 m_\odot$ in purple. Of the same $x$ and $m_\mathrm{bin}$, three points for $q_\mathrm{bin}=$1, 1/5 and 1/25 are plotted.}
\label{fig-a-tm-bocc}
\end{figure}

In the simulations carried out above, we have recorded, for each set of runs, the largest impact parameter allowing for each outcome to occur $b_\mathrm{occ}$. In Figure \ref{fig-a-tm-bocc}, we show in the top panel, $b_\mathrm{occ}$ normalised by $a_\mathrm{bin}$ for Jupiter's ejection. Only those for $v_\mathrm{inf}=1$ km/s are shown. For each $m_\mathrm{bin}$ and $a_\mathrm{bin}$, three points for $q_\mathrm{bin}=$1, 1/5 and 1/25 are plotted but these are close together and cannot be distinguished. Hence, the binary mass ratio only plays a minor role here.

Much as expected, $b_\mathrm{occ}/a_\mathrm{bin}$ depends on the binary mass positively -- more massive binaries are more capable of breaking up the SJ-pair. Then, $b_\mathrm{occ}/a_\mathrm{bin}$ is a monotonic decreasing function of $a_\mathrm{bin}$. We note when $a_\mathrm{bin}$ is small it may not be a proper normalisation factor. But this makes it easier to compare our result with the literature: for very tight binaries at $a_\mathrm{bin}=0.2$ au, $b_\mathrm{occ}/a_\mathrm{bin}$ can reach $\sim10^3$ and even at a few tens of au where the observed binary separation distribution peaks \citep[][and Section \ref{sec-mont}]{Raghavan2010}, still this ratio is several 10s. Thus, it seems that the value of 10 used by \citet{Li2015} might be not sufficient. For larger $a_\mathrm{bin}$, the ratio drops further. We note at $a_\mathrm{bin}=3125$ au, $\log b_\mathrm{occ}/a_\mathrm{bin}$ falls below 0 (i.e., $b_\mathrm{occ}/a_\mathrm{bin}<1$) for $m_\mathrm{bin}=1m_\odot$. This is apparently incorrect and can be probably attributed to the small chance for a component of a wide binary to interact with the SJ-pair at a close distance so as to eject Jupiter.

In this work, we are not only interested in Jupiter's ejection but also, in a general sense, situations where the status of the SJ-pair changes (SJ-change). This includes Jupiter's ejection, collision, capture by other stars as well as the exchange of the SJ-pair as a whole into a binary; see Figure \ref{fig-illu}. The bottom panel of Figure \ref{fig-a-tm-bocc} shows $b_\mathrm{occ}$ for SJ-change normalised now with respect to $a_\mathrm{tot}=a_\mathrm{SJ}+a_\mathrm{bin}$. The consideration is that when $a_\mathrm{bin}\ll a_\mathrm{SJ}$, the binary behaves much like a single star and then it is Jupiter's ejection and capture by the binary that dominate; so now it is $a_\mathrm{SJ}$ that is more relevant. But when $a_\mathrm{bin}\gg a_\mathrm{SJ}$, probably the exchange of the SJ-pair into a binary becomes more frequent and in this case, $a_\mathrm{bin}$ is a more appropriate measure. By using the normalisation factor $a_\mathrm{tot}$, both two cases are taken account of. In general, $b_\mathrm{occ}/a_\mathrm{bin}$ is positively dependent on $m_\mathrm{bin}$ but negatively on $a_\mathrm{tot}$. And again, we observe $\log b_\mathrm{occ}/a_\mathrm{tot}$ falls below 0 (i.e., $b_\mathrm{occ}/a_\mathrm{tot}<1$) at $a_\mathrm{tot}=(3225+5)$ au, a result of our limited number of runs.

In this section, we have iteratively increased $b_\mathrm{max}$ so it is large enough for binaries with parameters picked from a grid. However, when the binary parameter is drawn from a continuous distribution (Section \ref{sec-mont}), such an approach cannot be applied. Thus we want to use $b_\mathrm{occ}$ recorded here to shed light on $b_\mathrm{max}$ to be used in Section \ref{sec-mont} by deriving a parameter dependence for $b_\mathrm{occ}/a_\mathrm{tot}$. The starting assumption is that only encounters achieving a certain encounter distance $r_\mathrm{enc}$ can give rise to the outcomes of interest and this distance is the sum of the semimajor axes of the two pairs times a factor $r_\mathrm{enc}=F \,a_\mathrm{tot}$. Then according to gravitational focusing \eqref{eq-bmax-1}, the impact parameter corresponding to this encounter distance is
\begin{equation}
{b_\mathrm{occ}\over a_\mathrm{tot}}=F \sqrt{2Gm_\mathrm{tot}\over a_\mathrm{tot}v^2_\mathrm{inf}}\sqrt{1+{a_\mathrm{tot}v^2_\mathrm{inf}\over 2Gm_\mathrm{tot}}}
\end{equation}
where $m_\mathrm{tot}=m_\mathrm{SJ} +m_\mathrm{bin}$. In the case where $v_\mathrm{inf}$ is small compared to $\sqrt{2Gm_\mathrm{tot}/a_\mathrm{tot}}$, the escape velocity between the two pairs when they are $a_\mathrm{tot}$ apart, we can drop the second square root on the right hand side. Taking the logarithm of both sides of the equation, the result reads
\begin{equation}
\label{eq-bocc}
\log {b_\mathrm{occ}\over a_\mathrm{tot}}=\log F+{1\over2}\log m_\mathrm{tot}-{1\over2}\log a_\mathrm{tot}-\log v_\mathrm{inf}.
\end{equation}
This clearly explains the overall dependence of $b_\mathrm{occ}/a_\mathrm{tot}$ on $m_\mathrm{tot}$ and $a_\mathrm{tot}$ as observed in the bottom panel of Figure \ref{fig-a-tm-bocc}.

\begin{figure}
\includegraphics[width=\columnwidth]{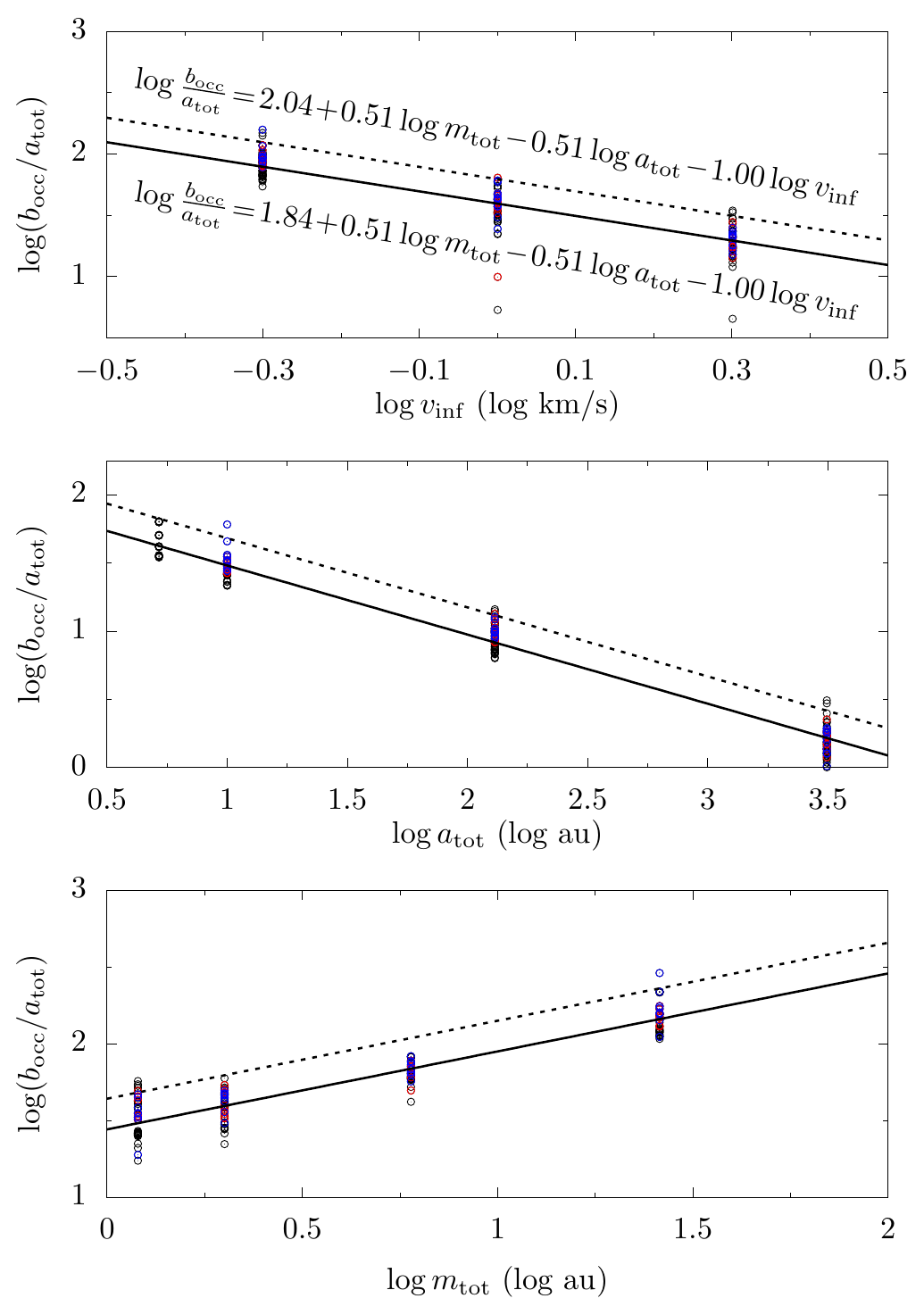}
\caption{The largest impact parameter $b_\mathrm{occ}$ observed for the status of SJ-pair to be changed in the scattering. In the three panels, $b_\mathrm{occ}$ is normalised according to Equation \eqref{eq-occc-fit} such that the binary parameters are $m_\mathrm{bin}=1 m_\odot$, $a_\mathrm{bin}=1$ au and $v_\mathrm{inf}=1$ km/s where applicable. Black points show simulations with $e_\mathrm{bin}=0$, red for $e_\mathrm{bin}=0.5$ and blue for $e_\mathrm{bin}=0.99$. The solid line represents a fit to the black points \eqref{eq-occc-fit} (see text for details) and the dashed line is additionally shifted vertically by 0.2 dex.}
\label{fig-fitted-bocc}
\end{figure}

Inspired by this, we vary the constants and fit $b_\mathrm{occ}$ using a functional form of $\log {b_\mathrm{occ}\over a_\mathrm{tot}}=c_0+c_1\log m_\mathrm{tot}+c_2\log a_\mathrm{tot}+c_3\log v_\mathrm{inf}$ and $b_\mathrm{occ}$ is averaged over the binary mass ratios. However, as discussed for Figure \ref{fig-a-tm-bocc}, at $a_\mathrm{tot}=(3125+5)$ au, $b_\mathrm{occ}$ falls below $a_\mathrm{tot}$, which is an artefact of our perhaps insufficient number of runs. Hence, in our fit, we let $b_\mathrm{occ}=a_\mathrm{tot}$ if $b_\mathrm{occ}<a_\mathrm{tot}$. A least square fit gives 
\begin{equation}
\label{eq-occc-fit}
\log {b_\mathrm{occ}\over a_\mathrm{tot}}=1.84+0.51\log m_\mathrm{tot}-0.51\log a_\mathrm{tot}-1.00\log v_\mathrm{inf}.
\end{equation}
Here the length is measured in au, mass in the solar mass, and velocity in km/s. The fitted coefficients differ from the model \eqref{eq-bocc} by at most a few per cent. In Figure \ref{fig-fitted-bocc}, we show the fit (solid line) compared to the data normalised in such a way that the encountering binary is of $m_\mathrm{bin}=1 m_\odot$, $a_\mathrm{bin}=1$ au and $v_\mathrm{inf}=1$ km/s whenever possible. Though the dispersion is a few times 0.1 dex, the overall trends are well reproduced. The fit will instruct us in Section \ref{sec-mont} in choosing $b_\mathrm{max}$.

Finally, we make a brief comment on the binary eccentricity. This parameter has not been extensively surveyed in our simulations. According to \citet{Hut1983a,Heggie1996}, it is the binary mean relative velocity that matters but not the instantaneous velocity, and the dependence of exchange and ionisation on eccentricity is weak or non-existent \citep{Antognini2016}. Here as in Table \ref{tab-param}, we have carried another two sets of simulations with $e_\mathrm{bin}=0.5$ and 0.99 for binaries of $q_\mathrm{bin}=1$ and $a_\mathrm{bin}\ge5$ au, the resulting $b_\mathrm{occ}$ are shown as red and blue points in Figure \ref{fig-fitted-bocc}. As expected, both roughly fall within the dispersion already spanned by the $e_\mathrm{bin}=0$ simulations.

\subsection{Cross-sections for the ejection of Jupiter}
The cross section of an event can be calculated using Equation \eqref{eq-cross-def}. As discussed before, the dynamics of the four body encounter is complex and numerous outcomes could result. Here we only discuss the cross section for the ejection of Jupiter; in Section \ref{sec-mont}, we will study in more detail different outcomes arising from Monte-Carlo simulations.

\begin{figure*}
\includegraphics[width=1.4\columnwidth]{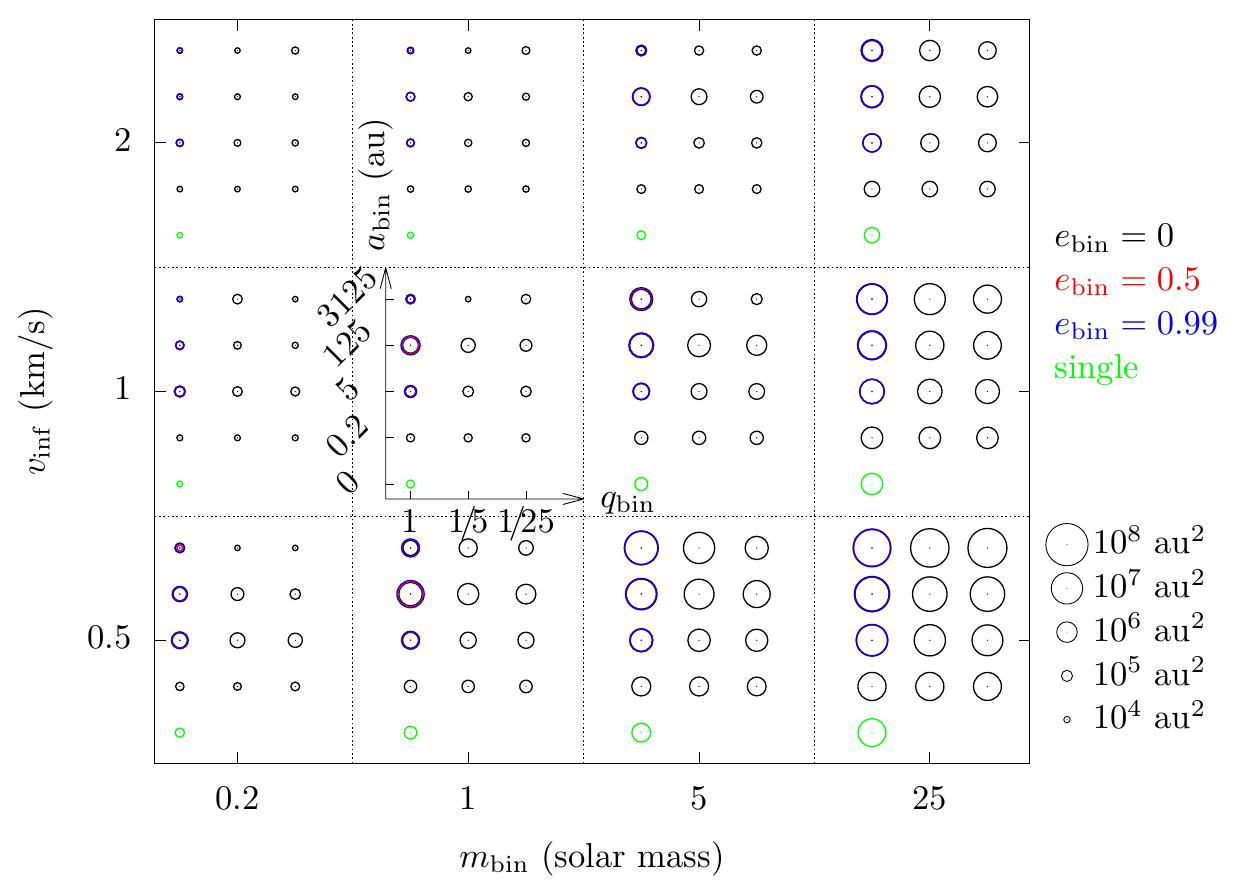}
\caption{Cross section for the ejection of Jupiter for scattering with a binary or a single star. The $x$-axis is the binary mass $m_\mathrm{bin}$ and $y$-axis the velocity at infinity $v_\mathrm{inf}$ between the SJ-pair and the binary. In each subplot, as bordered by dotted lines, the points have the same $m_\mathrm{bin}$ and $v_\mathrm{inf}$ but different binary mass ratio $q_\mathrm{bin}$ (horizontal direction) and binary separation $a_\mathrm{bin}$ (vertical direction). The points' sizes have to do with the size of the cross section, correlation shown on the bottom right. Black points are for binary eccentricity  $e_\mathrm{bin}=0$, red for $e_\mathrm{bin}=0.5$ and blue for $e_\mathrm{bin}=0.99$. Green points show results for the SJ-pair encountering a single star marked at $a_\mathrm{bin}=0$ au.}
\label{fig-tm-mr-v-a}
\end{figure*}

In Figure \ref{fig-tm-mr-v-a}, we use circles to show the cross section for Jupiter's ejection. There, the large $x$-axis is the binary total mass $m_\mathrm{bin}$ and the large $y$-axis is the velocity at infinity $v_\mathrm{inf}$. The plot is then divided into $4\times3=12$ subplots delimited by dotted lines, each of the same $m_\mathrm{bin}$ and $v_\mathrm{inf}$. In each of the subplot, the small $x$-axis shows the binary mass ratio $q_\mathrm{bin}$ and the small $y$-axis the binary semimajor axis $a_\mathrm{bin}$. Results for scatterings with binaries of zero eccentricity are shown in black, over-plotted with those with $e_\mathrm{bin}=0.5$ and 0.99 in red and blue and now only for $q=1$ and $a_\mathrm{bin}\ge 5$ au. Finally, we use the green circles to show the results for encounters with a single star, positioned at $a_\mathrm{bin}=0$ au.

Before discussing the trends, we first notice that the differences between black, red and blue circles are relatively small and the circles of the three colours almost overlap. Therefore, the binary eccentricity does affect much its ability to eject the planet from the SJ-pair \citep[cf.][]{Antognini2016}. Also, a tight binary behaves much like a single star since the cross section for a single star as marked at $a_\mathrm{bin}=0$ au (green) is close to that of a binary of $a_\mathrm{bin}=0.2$ au.

Overall, the cross sections span several orders of magnitude from $10^4$ to $10^8$ au$^2$ and $v_\mathrm{inf}$ and $m_\mathrm{bin}$ bear the clearest and steepest dependence. That on $v_\mathrm{inf}$ is apparent: when $v_\mathrm{inf}$ is small, gravitational focusing is strong and the cross section should be inversely proportional to $v^2_\mathrm{inf}$ \citep[e.g.,][]{Hut1983}. The dependence on $m_\mathrm{bin}$ is intuitive: the larger the intruder mass $m_\mathrm{bin}$, the more capable it is to eject Jupiter.

The dependence on $q_\mathrm{bin}$ is not straightforward and no clear trend exists. For encounters between a binary star and a single star with small encounter velocities, semianalytical scaling laws for the cross section of exchange were derived \citep{Heggie1996} and the dependence on the masses of the objects were unintuitive \citep[also][]{Fregeau2006}. From Figure \ref{fig-tm-mr-v-a}, it seems that if $m_\mathrm{bin}\le 5m_\odot$, equal mass binaries are more effective than ones with the same total mass but smaller mass ratios. As for $a_\mathrm{bin}$, we have discussed earlier that when this quantity is small, the binary can be treated as a single object, irrespective of $q_\mathrm{bin}$. In general, wider binaries are more effective in ejecting Jupiter. And thus binaries typically have larger cross sections than singles. But for very wide binaries $a_\mathrm{bin}=3125$ au and for high $v_\mathrm{inf}$, the effectiveness drops, since now the two components, with small revolution velocities, behave much like two singles.

\section{Monte carlo simulations}\label{sec-mont}
Above we have derived the cross section for the ejection of Jupiter for encountering a binary with parameters picked from a grid. Now in this section, we estimate the quantity for a realistic binary population and for various outcomes, not just the ejection of the planet.
\subsection{Simulation setup}
We first describe how the binaries are created. For each of them, we first draw the masses of the two components independently from a power-law distribution \citep{Kroupa2001} and in the range of 0.1 to 10 solar masses \citep{Li2015}. Then the physical radii of the two are calculated as in the previous section.

The orbital properties of binaries in open clusters are not well constrained \citep{Sollima2010} and we resort to those in the field. The semimajor axis and eccentricity are generated following \citet{Raghavan2010}: we first draw a binary orbital period $P_\mathrm{bin}$ from a lognormal distribution and then translate it into the semimajor axis $a_\mathrm{bin}$ with the above-generated masses. We restrict ourselves to pairs with $a_\mathrm{bin}$ larger than the sum of stars' radii. An upper limit of $a_\mathrm{bin}=10^4$ au is set because otherwise the binaries would be prone to encounter disruption in a cluster \citep[for instance][]{Kroupa1995,Parker2009} and other background stars may interlope the encounter we study \citep{Geller2015}; in addition, wide binaries also make the integration more CPU time-consuming. We will briefly discuss the implications of a smaller upper limit $a_\mathrm{bin}=1000$ au later.

The eccentricity of the binary $e_\mathrm{bin}$ is a function of $a_\mathrm{bin}$ or in a similar sense, $P_\mathrm{bin}$. For tight binaries with $P_\mathrm{bin}<12$ days, we let $e_\mathrm{bin}=0$ because of tidal circularisation \citep{Raghavan2010}. Then for wider binaries, it seems that the distribution is flat when $e_\mathrm{bin}<0.6$ \citep[\citealt{Raghavan2010}, but see also][]{Duquennoy1991}; that for $e_\mathrm{bin}>0.6$ is more poorly constrained and we make it also flat for simplicity.

We note that our binary population is not meant to be primordial \citep[for instance,][]{Kroupa1995}. For example, very tight binaries may have formed through a combination of Kozai--Lidov mechanism and tidal dissipation \citep[e.g.,][]{Fabrycky2007} and is thus not primordial. Also, in a broader sense, a more primitive distribution for $a_\mathrm{bin}$ could have more weight on the wide side that the observed one \citep{Kroupa1995}. But for the purpose of creating a reasonable representation of the binary population in an open cluster, our approach suffices \citep[e.g.,][]{Parker2009}.

After pinning down the binary parameters, we now proceed to discuss the encounter setup.  The typical velocity dispersion of an open cluster is $v_\mathrm{disp}\sim 1$ km/s \citep{Binney2008}. Then the relative velocity at infinity between two encountering stars would be $v_\mathrm{inf}\sim\sqrt{2}v_\mathrm{disp}$, also $\sim 1$ km/s. Here we simply draw $v_\mathrm{inf}$ from a Maxwellian distribution with a mean of 1 km/s. In Section \ref{sec-bocc}, we have shown that, the largest impact parameter $b_\mathrm{occ}$ observed in the simulation where status of the SJ-pair changes can be roughly fitted as a power law function of $m_\mathrm{bin}$, $a_\mathrm{bin}$ and $v_\mathrm{inf}$ but the scattering could be a few times 0.1 dex; see Figure \ref{fig-fitted-bocc}. Here we want to use that fit to provide an upper limit for the impact parameter, $b_\mathrm{max}$, that is large enough so no event of interest is missed and at the same time not too large, otherwise a large fraction of the simulation would not change the SJ-pair's status. Therefore we simply shift the fitted line vertically by 0.2 dex, or equivalently, increase the limit by 60\%. This is shown as the dashed lines in that figure. By doing so, we have covered the vast majority of the cases (most points are below the dashed lines) and only at the high ends of $a_\mathrm{bin}$ and $v_\mathrm{inf}$ and the lower end of $m_\mathrm{bin}$ do we possibly miss out scatterings that may modify the status of the SJ-pair. A total of $2.5\times10^7$ scattering runs for this set of simulations are performed.

The cross sections cannot be calculated directly as in Equation \eqref{eq-cross-def} because we do not have a universal $b_\mathrm{max}$. Instead, suppose we have obtained the cross section $\sigma_\mathrm{X}$ for an event X at some velocity $v$, the rate of X actually happening is then
\begin{equation}
\Gamma=\int n_v \sigma_\mathrm{X} v \,\mathrm{d}v,
\end{equation}
where $n_v$ is the number density of binaries with $v_\mathrm{inf}\in(v,v+\mathrm{d}v)$ in the cluster of interest. To remove the dependence on $v$, we opt to calculate the velocity-averaged cross section \citep[see also][]{Li2015}
\begin{equation}
\label{eq-cross-vel}
\langle\sigma_\mathrm{X}\rangle={1\over 1 \text{km/s}}\sum_{m_\mathrm{bin}} \sum_{a_\mathrm{bin}} \sum_{v_\mathrm{inf}}\pi b^2_\mathrm{max} {N_\mathrm{X}\over N} v_\mathrm{inf}.
\end{equation}
Here $b_\mathrm{max}$ is as described above a function of the $m_\mathrm{bin}$, $a_\mathrm{bin}$ and $v_\mathrm{inf}$. Then the rate of X occurring in an actual cluster becomes
\begin{equation}
\label{eq-occ}
\Gamma_\mathrm{X}=n_\mathrm{bin} \langle\sigma_\mathrm{X}\rangle \times 1 \, \mathrm{km/s}=2.4\times10^{-8}{n_\mathrm{bin}\over10 \, \mathrm{pc}^{-3}}{\langle\sigma_\mathrm{X}\rangle\over\mathrm{au}^2} {1\over100\,\mathrm{Myr}},
\end{equation}
where $n_\mathrm{bin}$ is the number density of binaries in that cluster. We note in the same cluster environment, the chance for X to occur is proportional to $\Gamma_\mathrm{x}$. This means that the rates for different outcomes in the same cluster can be compared in a relative sense using $\langle\sigma_\mathrm{X}\rangle$ without worrying about the detailed cluster properties. If we further that the number density is a few 10s pc$^{-3}$ and that the lifetime of the cluster is $T_\mathrm{c}=100$ Myr, the chance for X to happen is then 
\begin{equation}
\label{eq-rate}
p_\mathrm{X}\sim\Gamma_\mathrm{X}T_\mathrm{c}\sim2.4\times10^{-8}{n_\mathrm{bin}\over10 \, \mathrm{pc}^{-3}} {T_\mathrm{c}\over100\,\mathrm{Myr}} {\langle\sigma_\mathrm{X}\rangle\over\mathrm{au}^2}\sim10^{-7}{\langle\sigma_\mathrm{X}\rangle\over\mathrm{au}^2}.
\end{equation}

In \citetalias{Li2019,Li2020a}, we calculated the cross section for the ejection of a planet due to the flyby of a solar-mass star and we found that in general, our values were larger than that of \citet{Li2015} by about an order of magnitude (though our intruding star is more massive on average by a factor of a few).  As argued earlier in Section \ref{sec-grid}, we suspect that those authors might have underestimated the cross sections for scatterings between a planetary system and a binary. The same could be true for encounters involving a single star as their two sets of simulations for binaries/singles were carried out in similar ways \citep{Li2015}. To this end, we introduce another set of scattering simulations between the SJ-pair and a single star. The mass of the single star and the relative velocity at infinity are generated as before and $b_\mathrm{max}$ is chosen by letting $a_\mathrm{bin}=0$ au. A total of $10^6$ runs are performed for this set of simulations.

\subsection{Outcome classification}

 As briefly touched upon for the grid-simulations, the outcomes of scatterings between a binary and a planetary system are extremely rich and phase spaces inaccessible to scatterings between objects of similar masses, are now open \citep[e.g.,][]{Fregeau2006,Wang2020}. Here, we differentiate all four objects in our simulations and the permutations lead to a wealth of outcomes. Take the outcomes that have exactly one two-body collision for example. The collision may involve any two out of the four, giving rise to six possibilities and we are left with three objects (one is a merger). With three objects, the outcomes could be total ionisation (1 possibility), a binary plus an unbound single (3 possibilities) and a hierarchical three-body system (3 possibilities). Going through these permutations, $6\times(1+3+3)=42$ outcomes are in principle possible; 37 actually appear in our simulations. In total, more than 80 outcomes are observed. Presenting the cross section for each and every of them is prohibitive/unnecessary and classification is needed. Here we categorise the outcomes based on the behaviour of the planet.

\begin{figure}
\includegraphics[width=\columnwidth]{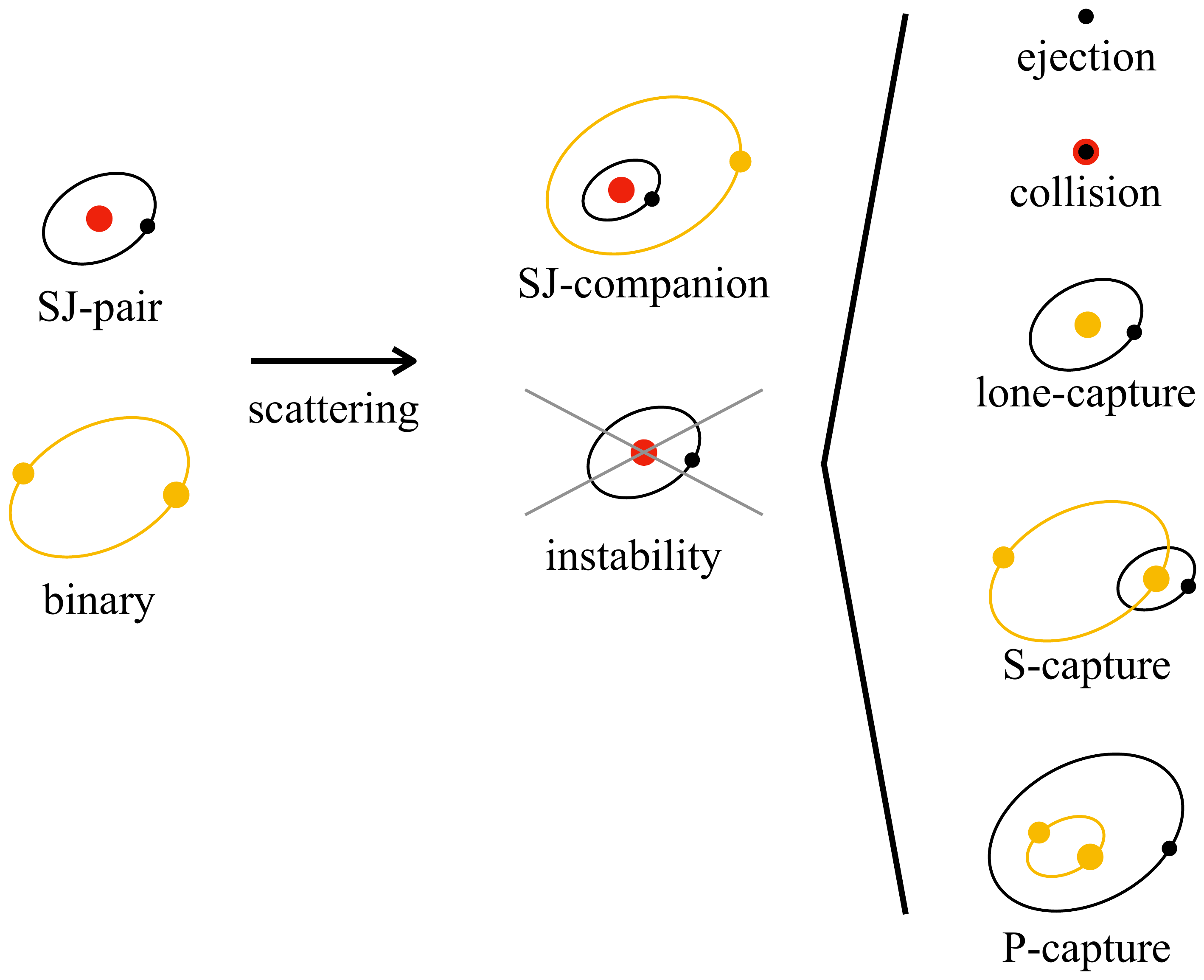}
\caption{Illustration of our classification scheme for the outcomes of the scattering between the SJ-pair and a binary. Here we only discuss the status of the SJ-pair. It can be broadly divided into two cases. The first, SJ-companion, is relevant when the SJ-pair is intact but accompanied by another star. In the second, instability, Jupiter is not revolving around the Sun. The second consists of five sub-cases: 1. ejection: Jupiter is not orbiting any star; 2. collision: Jupiter collides with a star; 3. lone-capture: Jupiter is captured by a single star without a tertiary; 4. S-capture: Jupiter is orbiting a component of a stellar binary; 5. P-capture: Jupiter is moving around a binary.}
\label{fig-illu}
\end{figure}

Our scheme contains seven categories. In the first, we have Jupiter as a free-floating planet without a host star; we call this ejection. In the second, Jupiter has collided with another object; we call this collision. In the third, Jupiter is orbiting another lone star (this star is not the Sun but could be a merger to which the Sun contributes via collision) and the new planetary system is not part of a hierarchical system; we call this lone-capture. In the fourth, Jupiter is revolving around another star and the pair is accompanied by a third object on a wider orbit; we call this S-capture (capture onto S-type orbits). In the fifth, Jupiter orbits around not a single star but a binary; we call this P-capture (capture onto P-type orbits). Then in the sixth, we include all cases where Jupiter is unstable from the Sun, i.e., all covered so far; we call this instability. Our seventh category is then applicable where the SJ-pair is intact and is, as a whole, exchanged into a wider binary as a component; so the configuration is SJ-companion. An illustration of the classification scheme can be found in Figure \ref{fig-illu}. As per Section \ref{sec-grid}, SJ-companion and instability can be called together as SJ-change.

\subsection{Cross sections and occurrence rates}\label{sec-sec-occ}
The velocity-averaged cross sections $\langle\sigma\rangle$ \eqref{eq-cross-vel} for the 7 types of outcomes are presented in Table \ref{tab-cross}. In the first column, we list the status and in the second and the third, the fourth, we show the corresponding $\langle\sigma\rangle$ for encountering a stellar population with only binaries (binary fraction $f_\mathrm{bin}=1$), only singles ($f_\mathrm{bin}=0$) and with half binaries and high singles ($f_\mathrm{bin}=0.5$; see below), respectively. Some outcomes are impossible for the single-only population and are not shown.

\begin{table*}
\centering
\caption{Velocity-averaged cross sections \eqref{eq-cross-vel} for different outcomes for scatterings between the SJ-pair and stellar populations of binaries or singles. In the first column, we have the status, in the second, we list the cross sections for encountering a population of binaries solely ($\langle\sigma\rangle_\mathrm{bin}$, hence the binary fraction is $f_\mathrm{bin}=1$), in the third for singles only ($\langle\sigma\rangle_\mathrm{sin}$, $f_\mathrm{bin}=0$), and in the fourth for half binaries and half singles ($\langle\sigma\rangle_\mathrm{eff}$, $f_\mathrm{bin}=0.5$). From the second to the eighth row, we show the cross sections for Jupiter's ejection, collision, capture by a lone star (lone-capture), capture onto S-type orbits (S-capture), capture onto P-type orbits (P-capture), unstable from the Sun (all 5 above and referred to as instability), and the exchange of the SJ-pair as a whole into a binary (forming a SJ-companion configuration); an illustration of the classification scheme can be found in Figure \ref{fig-illu}. Then the ninth and the tenth lines show those for the SJ-companion configuration and where KL timescale is shorter than binary disruption timescale, for stellar densities $n$=10/pc$^3$ and 100/pc$^3$, respectively.}
\label{tab-cross}
\begin{tabular}{c c c c}
\hline
&$\langle\sigma\rangle_\mathrm{bin}$ ($10^3$ au$^2$, $f_\mathrm{bin}=1$)&$\langle\sigma\rangle_\mathrm{sin}$ ($10^3$ au$^2$, $f_\mathrm{bin}=0$)&$\langle\sigma\rangle_\mathrm{eff}$ ($10^3$ au$^2$, $f_\mathrm{bin}=0.5$)\\
\hline
ejection&$168\pm1$&$25.9\pm0.2$&97.0$\pm$0.5\\
collision&$17.1\pm0.2$&$0.165\pm0.012$&8.63$\pm$0.15 \\
lone-capture&$5.43\pm0.15$&$8.85\pm0.10$&$7.14\pm0.14$ \\
S-capture&$4.62\pm0.24$&- &$2.31\pm0.12$\\
P-capture&$4.25\pm0.07$&- &$2.13\pm0.03$\\
instability&$200\pm1$&$35.0\pm0.2$&$118\pm1$\\
SJ-companion&$1640\pm3$&- &$820\pm2$\\
\hline
KL-excitation ($n$=10/pc$^3$)&$791\pm2$&-&$396\pm1$\\
KL-excitation ($n$=100/pc$^3$)&$575\pm1$&-&$288\pm1$\\
\hline
\end{tabular}
\end{table*}

Reading from the table, the cross-section for ejection is $1.7\times10^5$ au$^2$. In a very much similar setup, \citet{Li2015} reported a value smaller than ours by an order of magnitude. The cause, as analysed before, is probably that those authors did not sample impact parameters sufficiently large. Recently, \citet{Wang2020} calculated the cross sectional area of the ejection of a planet at 1 au from its host star for encountering binaries. Considering that the cross section should scale linearly with the planetary semimajor axis \citep[e.g.,][]{Heggie1996,Fregeau2006}, the result of \citet{Wang2020} would suggest an area of a few time $10^5$ au$^2$ for Jupiter, consistent with ours.

Our cross section for ejection for encountering singles is $2.6\times10^4$ au$^2$, larger than that of \citet{Li2015} by a factor of several. And again, the value obtained by \citet{Wang2020} was in rough agreement with ours.

The ratio between the cross sectional areas for ejection for encountering binaries and singles is about 6.6 in our work, compatible with 6 as reported in \citet{Antognini2016}. This ratio was measured as 3.6 in \citet{Li2015}.

Then the cross section for collision $\sigma_\mathrm{col}$ is then $1.7\times10^4$ au$^2$ for binaries and only 170 for singles, both in agreement with \citet{Wang2020} though their values were measured for a planet at 1 au. We can refer to \citep{Fregeau2004} for a possible explanation for the perhaps-surprising coincidence between our result and that of \citet{Wang2020}. For encounters between a stellar binary and a single star, table 4 of \citet{Fregeau2004} showed that when the binary separation $a_\mathrm{bin}$ is much larger than the stellar radii, $\sigma_\mathrm{col}$ scales roughly with $a_\mathrm{bin}v^2_\mathrm{crit}/v^2_\mathrm{inf}$. When applied to our simulations, $v_\mathrm{inf}$ can be dropped because it is generated irrespective of the binary properties. From Equation \eqref{eq-vcrit}, $v^2_\mathrm{crit}\propto 1/a_\mathrm{bin}$. Then the dependence of $\sigma_\mathrm{col}$ on $a_\mathrm{bin}$ is canceled. This could be the reason why the cross section for a planet at 5 au is not so different from one at 1 au.

 It has been shown by \citet{Wang2020} that compared to single stars, binaries significantly enhance the cross sections for both ejection and collision. However, the extents of boost in the two outcomes are not the same. While that for ejection is about a factor of several, that for collision is  two orders of magnitude. As a consequence, for encountering single stars, the ratio between planetary collision and ejection is $\sim0.01$, but it grows to $\sim 0.1$ for encountering binaries.

Then in the 4th, 5th and the 6th rows, we have the cross sections for Jupiter's capture by other stars. For encountering binaries, the three capture scenarios, lone-, S- and P-capture have similar areas. For encountering singles, S- and P-capture are impossible; that for lone-capture is larger than scatterings with binaries by 60\%. For encountering single stars, $\sigma$ for capture is about a third of that of ejection, compatible with \citetalias{Li2019}. The orbitals of S- and P-captures from encountering binaries will be discussed later.

In all 6 cases discussed above, Jupiter is not revolving around the Sun, i.e., unstable from the Sun. In the seventh row, we have the cross section for Jupiter's instability. Hence, ejection is the most dominant source of instability. And the binaries are more effective than singles by a factor of seven.

Finally, far more SJ-pairs are stable but they end up in an SJ-companion configuration with a cross section of $1.6\times10^6$ au$^2$, 10 times of that of Jupiter's ejection. As to be discussed below, many of those systems are characterised by wide companion orbit that can be created with large $a_\mathrm{bin}$ and $b$, a reason why $\sigma$ is large. This outcome is not possible for scatterings with single stars \citep[see also][]{Fregeau2006}.


The cross sections can be used as Equations \eqref{eq-occ} by supplying the cluster stellar number density $n$ and the return is the absolute occurrence rates for the respective outcomes. For instance, Equation \eqref{eq-rate} tells that in a cluster of binary fraction 1, the probability for the formation of SJ-companion configure is $\sim10\%$ and that for Jupiter's ejection is $\sim1\%$. Or, without knowing the cluster parameters, the cross sections in Table \ref{tab-cross} can be used in a relative manner and the relative occurrence rates result.

Because of scatterings with binaries, for every 1 free-floating planet thus created, 0.1 collides with the stars, another 0.1 is captured by other stars, possibly residing within a stellar binary and additionally, 10 would, together with their original host, be accompanied by a tertiary star. And owing to scatterings with single stars, for every 1 planet ejected, <0.01 collide with the stars and 0.3 are captured by the intruder.

In an actual cluster, these ratios will be further modulated by the binary/single ratio. Here we follow \citet{Li2015} to define an effective cross section
\begin{equation}
\label{eq-eff}
\langle\sigma\rangle_\mathrm{eff}=f_\mathrm{bin}\langle\sigma\rangle_\mathrm{bin}+(1-f_\mathrm{bin})\langle\sigma\rangle_\mathrm{sin},
\end{equation}
where $f_\mathrm{bin}$ is the binary fraction of the cluster and $\langle\sigma\rangle_\mathrm{bin}$ and $\langle\sigma\rangle_\mathrm{sin}$ the cross sections for scatterings with binaries and singles, respectively. As discussed in Section \ref{sec-intro} the binary fraction in open clusters is in general consistent with that of the field and the primordial value is probably higher \citep{Kroupa1995,Kroupa2001a,Parker2009,Marks2012}. For $f_\mathrm{bin}=0.5$, the resulting $\langle\sigma\rangle_\mathrm{eff}$ is listed in the fourth column of Table \ref{tab-cross}. While the absolute numbers may change considerably compared to binaries/singles only, the ratios between different outcomes do not change much and those argued for scatterings with binaries roughly hold.

\begin{figure*}
\includegraphics[width=1.4\columnwidth]{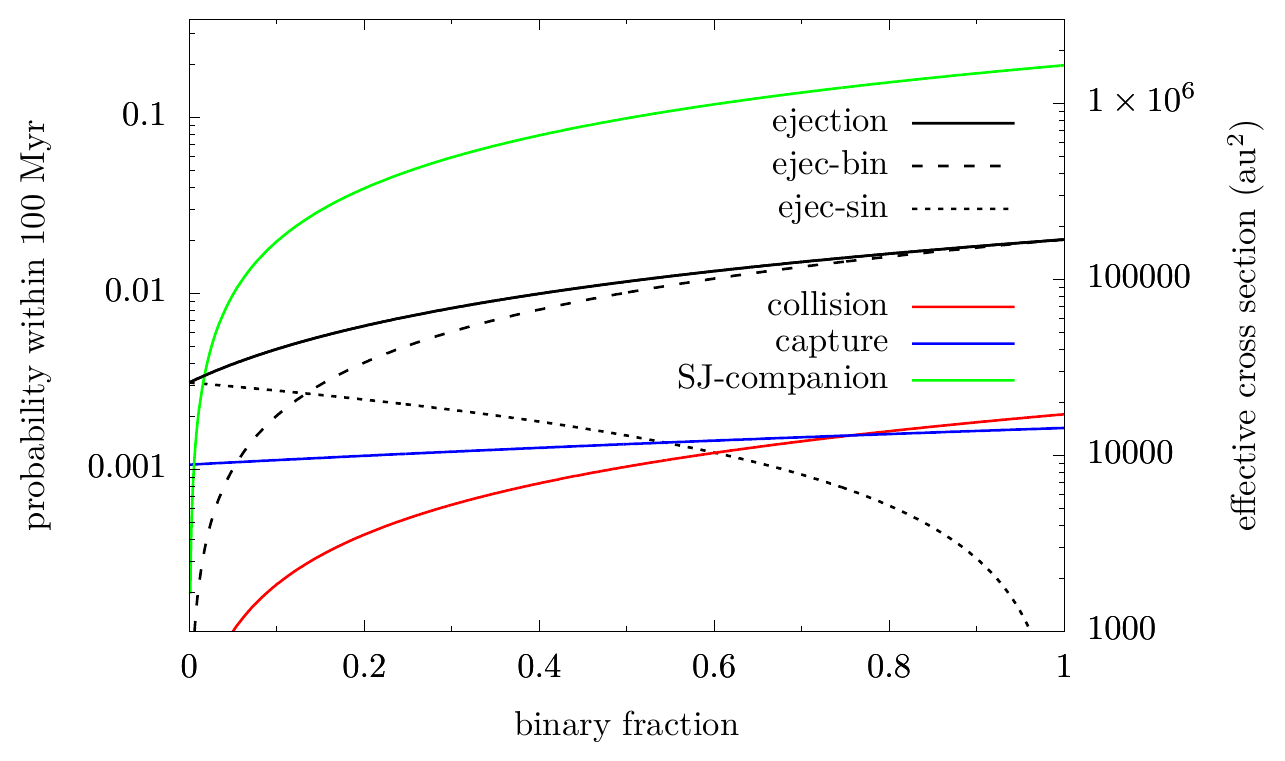}
\caption{The probabilities of various different events and their cross sections as a function of the binary fraction of the cluster. The solid black, red, blue, and green lines show Jupiter's ejection, collision, capture (including all three scenarios, Figure \ref{fig-illu}), and the formation of the SJ-companion configuration. The black dashed and dotted lines are a breakdown of ejection into the contribution of the binaries and singles, respectively. The right $y$-axis is the absolute cross section measured in au$^2$. The left $y$-axis is the probability of an SJ-pair to actually experience those outcomes within the lifetime of a fiducial cluster as estimated using Equation \eqref{eq-rate}. We assume the cluster has a stellar density of $n$=50 pc$^{-3}$ and a lifetime of 100 Myr.}
\label{fig-bin_vs_sin}
\end{figure*}

More generally in Figure \ref{fig-bin_vs_sin}, the corresponding probability of different events occurring during the lifetime of a cluster and the effective cross sections of them are shown as a function of the binary fraction of the cluster. Black lines are for Jupiter's ejection, red for collision, and blue for capture; the green line is for the formation of SJ-companion configuration. That for ejection has also been broken down into the contribution from the binaries (dashed line) and from the single stars (dotted line). Here, the cluster has, as assumed in Equation \eqref{eq-rate}, a stellar density \footnote{We note that here the stellar density is for the stellar systems: a single star and a binary star have the same contribution. This also clarifies how binary fraction is defined in this work: the number of binary systems divided by the sum of binary pairs and single stars, i.e., $f_\mathrm{bin}=N_\mathrm{bin}/(N_\mathrm{sin}+N_\mathrm{bin})$.} of $n=$50 pc$^{-3}$ and a lifetime of $T_\mathrm{c}=100$ Myr.

For the ejection of Jupiter, the probability increases from 0.003 for a cluster composed of only single stars to 0.02 in the dominance of binaries (solid black line). Notably, with a binary fraction of only $f_\mathrm{bin}\sim0.15$, the binaries are already as important as the single stars in ejecting Jupiter (dashed and dotted black lines cross at that binary fraction). The red line suggests that, at $f_\mathrm{bin}\sim0$, planetary collision is very unlikely with a probability $<10^{-4}$; this number soars to 0.02 at $f_\mathrm{bin}\sim 1$. From Equation \eqref{eq-eff}, we deduce that with a binary fraction of only $\sim$0.01, it is still the binaries that are more effective in causing the planet's collision. In contrast, the capture probability only increases mildly when binaries take up a larger fraction and is around 0.001. Finally, the formation of the SJ-companion configuration is apparently increasing with $f_\mathrm{bin}$ and is higher than that of ejection by an order of magnitude for $f_\mathrm{bin}\gtrsim0.2$. As will be seen later in Section \ref{sec-sjt}, for about half of these SJ-companion systems, the companion is able to excite the orbit of Jupiter through the Kozai--Lidov mechanism and we refer to those as being Kozai--Lidov damaged. Then from Figure \ref{fig-bin_vs_sin}, so long as the binary fraction is larger than a few times 0.01, the probability of exchanging into a binary (green line) is more than double the probability of an ejection of Jupiter as a direct result of a scattering (black line). Therefore as about half of exchanges into binaries lead to Kozai-Lidov excitation, damage to a planetary system through this mechanism will be more common than direct ejection.

A sanity check concludes this section. In clusters of $\sim2000$ stars, depending on the exact setup of the cluster, $N$-body simulations suggest that a fraction of 1\% (the dispersion can be an order of magnitude) of planets at 5 au around a G-type star are ejected in a few 100s of Myr \citep[][see also \citealt{Cai2017}]{Fujii2019}, in agreement with Figure \ref{fig-bin_vs_sin}. Then, for the formation of SJ-companion configuration, \citet{Malmberg2011}, in a simulation with an initial $f_\mathrm{bin}=0.2$, showed the fraction to be 7\%. This is again, consistent with out estimate in Figure \ref{fig-bin_vs_sin}.

 Following \citet{Li2015}, in the above simulations, the masses of the two components of the binary are drawn independently \citep{Kroupa1995}. Another widely adopted approach is to first draw a mass for the more massive star at random then another also randomly but requiring that it is less massive than that of the first \citep{Parker2009,Parker2014}. We have conducted a smaller set of simulations of $2\times10^6$ scattering runs where the binary masses are generated in accordance with the second method. The resulting cross section for ejection is $117\times10^3$ au$^2$, dropping by 30\% compared to the first approach as in Table \ref{tab-cross}; collision decreases by 25\%, turning into $12.8\times10^3$ au$^2$; capture of all three scenarios has a total cross section of $8.22\times10^3$ au$^2$, implying a dip of 43\%; Finally, that for the formation of SJ-companion configuration is now $1400\times10^3$ au$^2$ -- a change of $-15\%$. With these drops, still the binaries dominate ejection and collision as long as the binary fraction is $\gtrsim10\%$ as seen from Figure \ref{fig-bin_vs_sin}.

\subsection{Planets in binaries}
As we have demonstrated, the probability for the planet to end up in a stellar binary after the scattering, mostly around the original host with a far-out companion, is much higher than other outcomes and reaches 0.1 in our fiducial cluster. Here we devote this section to the properties of the so-formed systems. When Jupiter is revolving around one of the two binary star components, we call it circumstellar or on S-type orbit and when orbiting the whole binary we refer to it as circumbinary or on P-type orbit (Figure \ref{fig-illu}). The S-type orbits are dominated by the SJ-companion configuration and S-capture contributes little (Table \ref{tab-cross}).

Before analysing the distribution those planets, we want to note that our stellar population, be it binary or single, is created in accordance with the observations. There is an implicit assumption: the population only applies to a given volume, implying the same limit on the impact parameter $b_\mathrm{max}$ for all scatterings. But in our simulations, a binary is assigned a $b_\mathrm{max}$ using its properties and, more massive, tighter and faster binaries have smaller $b_\mathrm{max}$. This means that the raw data are biased towards those binaries since as their smaller $b_\mathrm{max}$ naturally leads to a higher interaction rate. Additionally, we want to tie also the distribution of the planets in binaries to the rate of occurring; then the encounter velocity $v_\mathrm{inf}$ needs to be taken into consideration. So when calculating the distribution, we have applied a debiasing factor of $b^2_\mathrm{max}v_\mathrm{inf}$ for each scattering experiment, the same as that for the calculation cross section \eqref{eq-cross-vel} (the effect of this debiasing can be seen in the top panel of Figure \ref{fig-eccinccdf} below). With this correction implemented, our distribution represents the chance of occurring for the corresponding system architectures.

\begin{figure}
\includegraphics[width=\columnwidth]{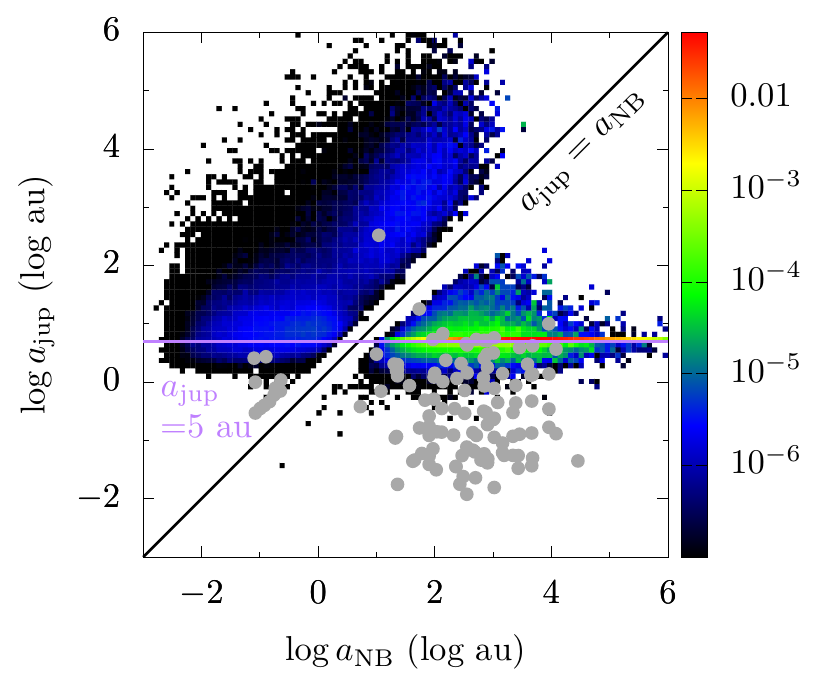}
\caption{Distribution of binary and planetary orbits when Jupiter is in a binary system. The $x$-axis is the binary semimajor semimajor axis $a_\mathrm{NB}$ and $y$ that of Jupiter $a_\mathrm{jup}$. Warmer colours represent higher density. The grey points are the observed planets in binary systems. The black line marks $a_\mathrm{jup}=a_\mathrm{NB}$ and the purple line the initial planetary semimajor axis $a_\mathrm{jup}=5$ au. The distributions have been debiased and normalised and are proportional to the rate of occurring for the respective $(a_\mathrm{NB},a_\mathrm{jup})$ pair.}
\label{fig-ater-ajup}
\end{figure}

In Figure \ref{fig-ater-ajup}, we show the distribution of the binary semimajor axis and that of Jupiter for S- and P-type configurations. Overall, both cover wide ranges spanning several orders of magnitude. There are two regions, one on the left above the black line with Jovian semimajor axis $a_\mathrm{jup}$ larger than the that of the stellar binary $a_\mathrm{NB}$ (meaning ``new binary'') for P-type orbits and the other on the right under the black line with $a_\mathrm{jup}<a_\mathrm{NB}$ for S-type orbits. In between, a void region where the planetary orbit is unstable \citep[e.g.,][]{Holman1999} exists. The observed population \citep{Schwarz2016a}\footnote{Retrieved on 2020 June 16 from \url{https://www.univie.ac.at/adg/schwarz/bincat_binary.html}. Only planets around main sequence stars are shown and only these with both values listed are shown.} has been over-plotted as grey points \citep[see also][]{Martin2018}. We note that the observed systems here do not represent a complete sample and are just a compilation of what has been detected. Known biases exist in the observed sample \citep[e.g.,][]{Eggenberger2010,Martin2018}.

For S-type orbits, perhaps the most prominent feature of the distribution from our scattering runs is a horizontal over-density strip with $a_\mathrm{jup}$ close to 5 au (the initial value and the purple horizontal line) and $a_\mathrm{NB}$ from 100s to 10s of 1000s of au, implying that the Jovian orbit is not disturbed much during the scattering \citep[also][]{Fragione2019}. Other than the strip, our planets are distributed between a few to a few 10s of au, much wider than observed population. This could be an artefact reflecting our lack of knowledge of wide-orbit planets. On the other hand, the binary separation of the observed population may seem to be not inconsistent with our results and this will be addressed later. We note that our binaries can be as wide as $10^6$ au and these cannot remain stable for long in a cluster \cite[e.g.,][and also see below]{Parker2009,Marks2012}. We count them anyway for completeness and for the fact that we are here only considering the scattering process itself but not the evolution afterwards.

Our P-type orbits are very much diffusive with two slight broad concentrations with binary separations tighter than 1 au or wider than 10 au. In contrast, the observed P-type planets mostly have very tight orbits just outside the stability limit where the binary separation is $<$ 1 au \citep{Martin2018}. One exception sticking out is the FW Tau system. There, the planet was directly imaged at about 330 au from a central binary separated by 11 au \citep{White2001,Kraus2013}; it sits comfortably in one of our marginally denser regions. Besides, a few other wide-orbit P-type planets are listed in the category by \citet{Schwarz2016a} but with no measured binary semimajor axis and are not shown. Among these, HD 106906 AB b \citep{Bailey2014} is 650 au from its host binary that is composed of two components sub-au apart \citep[see][and references therein]{Rodet2017}. \citet{Rodet2017} suggested that this planet could be on its way of ejection due to resonances, stabilised by a coincident stellar encounter. Here our simulations imply that such circumbinary wide orbit planet can also be created by direct capture by a binary, like that by a single star \citep[][\citetalias{Li2019}]{Mustill2016}.

Then in Figure \ref{fig-eccinccdf}, we show the cumulative distribution function (CDF) of the planet's and the binary's orbits, all debiased as described above. In the top panel, CDFs for inclination are presented. The red and blue lines represent those for the binary orbits, measured against the orbital norm of the planet, for P- and S-capture respectively. Both follow a sinusoidal function, suggestive of isotropic distribution. So the orbital angular momenta of the planet and the binary are not related during the brief scattering. But this non-correlation also implies that the so-formed systems tend to have the two orbital planes highly inclined with respect to each other, i.e., a preference for 90$^\circ$. Then a significant fraction of the S-type planets may be subject to large amplitude oscillations in the orbital eccentricity due to Kozai--Lidov mechanism \citep[e.g.,][]{Malmberg2007a,Antognini2016}. We will discuss this matter further later. The CDFs for the planets on P- and S-type orbits are shown in green and black, now measured against the direction of the planet's initial orbital angular momentum; this measures the relative change in the planet's orbital norm before and after the scattering. The fact that the distribution of the P-type orbits mimics a sinusoidal curve means that the capture process has randomised the planets' orbital planes, or that secular evolution could have taken place, adding further stochasticity \citep{Farago2010}. And noticeably, that for S-type orbits (black), on the contrary, favours small inclinations with over 70\% lower than 10$^\circ$, meaning that the formation of the SJ-companion configuration does not affect the motion of the planet much.

\begin{figure}
\includegraphics[width=\columnwidth]{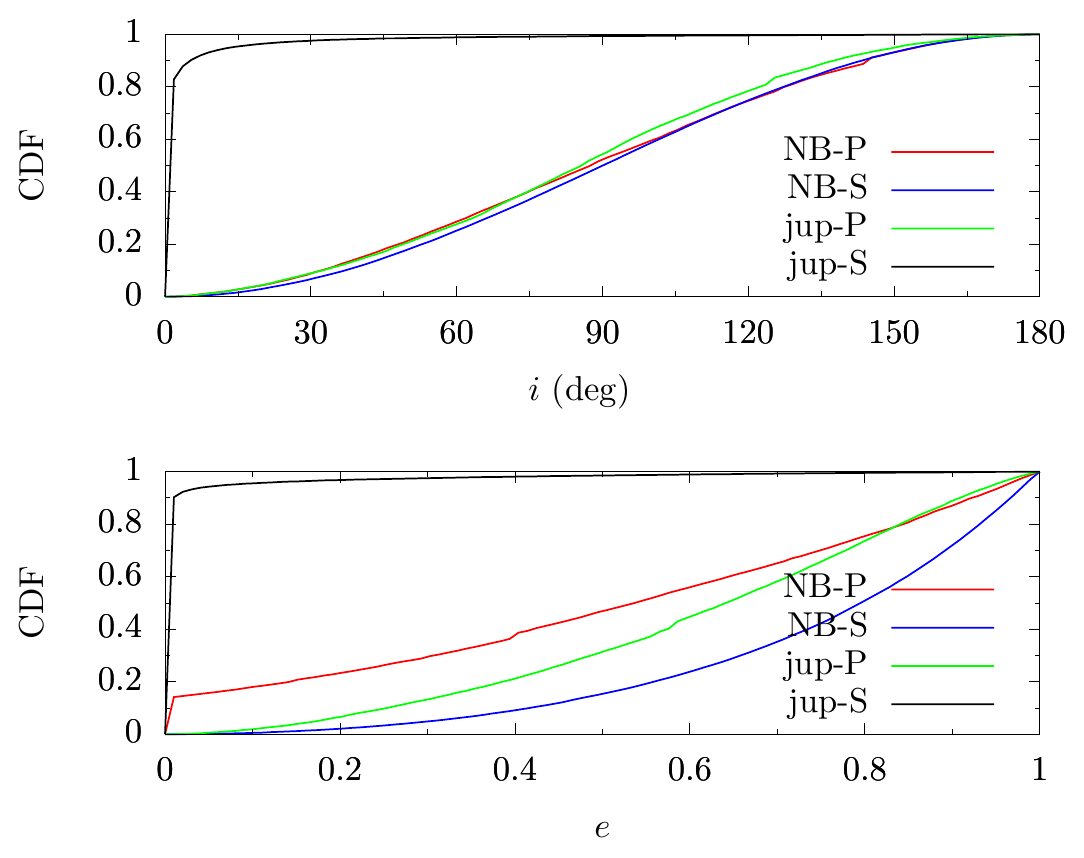}
\caption{Cumulative distribution function of the orbital elements of Jupiter and the binary star. In the top panel, we show the inclination: red and blue for the binary measured against the orbital angular momentum of Jupiter on P- and S-type orbits; green and black for Jupiter on P- and S-type orbits measured against the initial norm to Jupiter's orbit before the scattering. In the bottom, we show the distribution for eccentricity, colour scheme the same.}
\label{fig-eccinccdf}
\end{figure}

The CDFs for eccentricity are shown in the bottom panel of Figure \ref{fig-eccinccdf}. There is an obvious over-abundance of near circular orbits for the binaries hosting circumbinary planets (red), as a consequence of the capture of Jupiter by very tight binaries, to which we assign an eccentricity of zero. These binaries can keep their orbit unchanged during the scattering. On the other hand, that of the binary with a planet on S-type orbit (blue) is much hotter, close to a thermal distribution \citep[for example,][]{Antognini2016}. Then the CDF for the planets on P-type orbits (green) is close to thermal but colder, possibly as a result of the preferential removal of extreme eccentricities by the binary. The distribution for S-type planet's orbit (black) takes a different shape. Most planets retain cold orbits, with 80\% under an eccentricity of 0.1. This again can be attributed to the fact that the SJ-pair is often exchanged into a wide binary as a whole and the SJ relative orbit is not perturbed much \citep{Fragione2019}.

\subsection{Sun-Jupiter pair with a companion}\label{sec-sjt}

From now on, we concentrate on the systems of SJ-companion configuration because they are much more common than any other outcomes by at least an order of magnitude.

\begin{figure}
\includegraphics[width=\columnwidth]{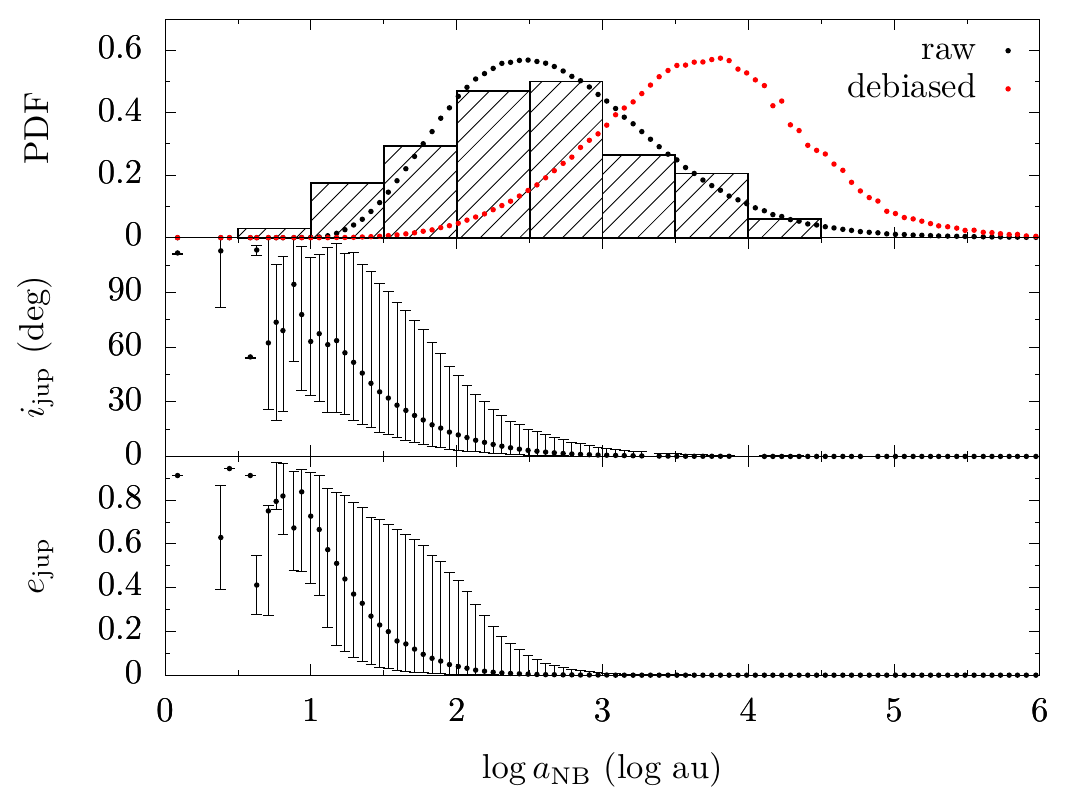}
\caption{Distribution of orbits of the SJ-companion systems. In the top panel, the probability density function of the companion semimajor axis $a_\mathrm{NB}$ is shown (points; black for raw data and red for debiased population) together with the observed separation between the two components of binaries hosting S-type exoplanets (boxes); they are normalised such that the area below is unity. In the middle and the bottom panels, we show the distribution of the Jovian inclination and eccentricity as a function of $a_\mathrm{NB}$: points for the median and error bars for the 16th and 84th percentiles.}
\label{fig-sjtejupijup}
\end{figure}

Intuitively the planets' properties should depend on the companion orbital separation $a_\mathrm{NB}$. In Figure \ref{fig-sjtejupijup} we show the Jovian orbit distribution as a function of the $a_\mathrm{NB}$. In the top panel, we first plot the probability density function (PDF) of $a_\mathrm{NB}$ using the points. Here the black points represent the raw data directly taken from all SJ-companion systems from the simulations whereas the red points show a population that has been debiased. As discussed before, the raw data is biased towards binaries of small impact parameters. Hence, before debiasing, most of the SJ-companion systems are only a few 100s of au wide whereas afterwards, the weight shift to over 1000s of au. Those from the observed sample shown in boxes; see \citet{Kraus2016} for a comparison between S-type planet-hosting binaries and normal field binaries.\footnote{About 110 such planets are detected around 70 stars, so many hosts have more than one planet. Here in Figure \ref{fig-sjtejupijup}, a star is counted once no matter how many planets it has.} As can be seen, the observations in general agree with our raw data but our debiased population has much wider orbits. However, we note that here we only care about the scattering itself where as later cluster evolution may shepherd the distribution towards the small end due to the breakup of very wide companions.

Then how are the planet's orbits affected by the scattering? In the middle and the bottom panels of Figure \ref{fig-sjtejupijup}, we show the distributions of Jupiter's inclination (measured against the initial planetary orbital plane) and eccentricity as a function of $a_\mathrm{NB}$, points showing the median and error bars marking the 16th and 84th percentiles. Even for $a_\mathrm{NB}\sim 100$ au, half of the planets are only slightly disturbed with eccentricities under 0.1 and inclinations under 10$^\circ$. For wider separations where most of the binaries do end up with, the orbital excitation of Jupiter is even smaller. Combined these with the top panel, we deduce that for most of so-formed SJ-companion systems, Jupiter's orbit is mostly not affected.

But this is not the entire story. In our simulations, we are only modelling the brief scattering process itself, whereas the long-term aftermath of the formation of the SJ-companion configuration may radically affect the planet's orbit. We here briefly discuss the Kozai--Lidov mechanism \citep{Kozai1962,Lidov1962}. A major phenomenon of this mechanism is a possible large-amplitude oscillation in the planet's orbital eccentricity driven by the companion; for a multi-planet system, the outmost planet can be highly excited, leading to the planets' orbital crossing and the system's instability \citep{Malmberg2007a}. This mechanism is most effective when the relative inclination between the planetary orbital plane and that of the companion is larger than 40$^\circ$ [$\in(40,^\circ,140^\circ)$]. The top panel of Figure \ref{fig-eccinccdf} shows that this angle (blue) agrees with an isotropic distribution for the orbital norms. Hence, the CDF of this inclination follows a sine function. Then the chance for it to fall between 40$^\circ$ and 140$^\circ$ is roughly 0.8 and thus the vast majority of the companions can excite Jupiter's orbit, given enough time.

However, even if the inclination is high enough, the Kozai--Lidov mechanism may still be suppressed by other effects that drive the planet's orbit to precess faster, for example, other planets \citep{Innanen1997} or relativistic effect \citep{Fabrycky2007}. Here we discuss another factor characteristic of the clustered environment -- the disruption of the SJ-companion configuration due to further scatterings. Future encounters between the SJ-companion and other objects may eject the companion via exchange or ionisation, terminating the Kozai--Lidov mechanism. Then we need to estimate the lifetime of the SJ-companion system $T_\mathrm{NB}$ and compare it with the timescale of the Kozai--Lidov mechanism $T_\mathrm{KL}$.

This $T_\mathrm{KL}$ depends on the binary separation $a_\mathrm{NB}$ sensitively to the third power \citep{Kiseleva1998}. The cross section to disrupt the SJ-companion is proportional to $a_\mathrm{NB}$ \citep{Hut1983a,Heggie1996} and so is the rate according to Equation \eqref{eq-occ}. Hence, the timescale $T_\mathrm{NB}$ is inversely dependent on $a_\mathrm{NB}$. Then, when $a_\mathrm{NB}$ is large, the binary can be broken before the Kozai--Lidov mechanism excites Jupiter's orbit.

The timescale $T_\mathrm{KL}$ can be readily evaluated as per \citet{Kiseleva1998}. However, the lifetime of the SJ-companion system in a cluster is not straightforward to estimate \citep[e.g.,][]{Parker2009}. Here we simply use the reciprocal of the occurrence rate \eqref{eq-occ} of its breakup as a proxy of the lifetime. Then the cross section of breakup $\sigma$ and the stellar density $n$ are needed. For the latter, we consider a population of singles only and discuss two situations $n=10$ or 100/pc$^3$. For the former, we have to perform a case-to-case analysis. If omitting the much less massive Jupiter, the SJ-companion system can be treated as a stellar binary and thus, encounters between that system and a stellar population of singles only can be thought of as that binary-single scatterings. If the relative velocity at infinity is smaller than a critical value \citep[][and see, Equation \eqref{eq-vcrit} for example]{Hut1983}, full ionisation of the three bodies is not possible and the only way to destroy the binary is via an exchange action (though a new binary is formed, potentially containing the SJ-pair again). Now the formalism in \citet{Heggie1996} applies. Or when the velocity is large, both ionisation and exchange are allowed and those in \citet{Hut1983a} should be used. We note both two prescriptions are in some sense (semi-) analytical asymptotic scaling laws and no rigour shall be assumed. For example, the so-evaluated $\sigma$ is not continuous at the critical velocity; also, strictly speaking, the formulae in \citet{Hut1983a} are relevant only for scattering between equal-mass stars, as required for a clean velocity exchange. Given these uncertainties, we opt to only consider the encountering single star to be 0.3 solar mass that is approaching the SJ-companion system at 1 km/s. Then the critical velocity is calculated and depending on whether this velocity is larger than 1 km/s or not, one of the two prescriptions discussed above applies.

\begin{figure}
\includegraphics[width=\columnwidth]{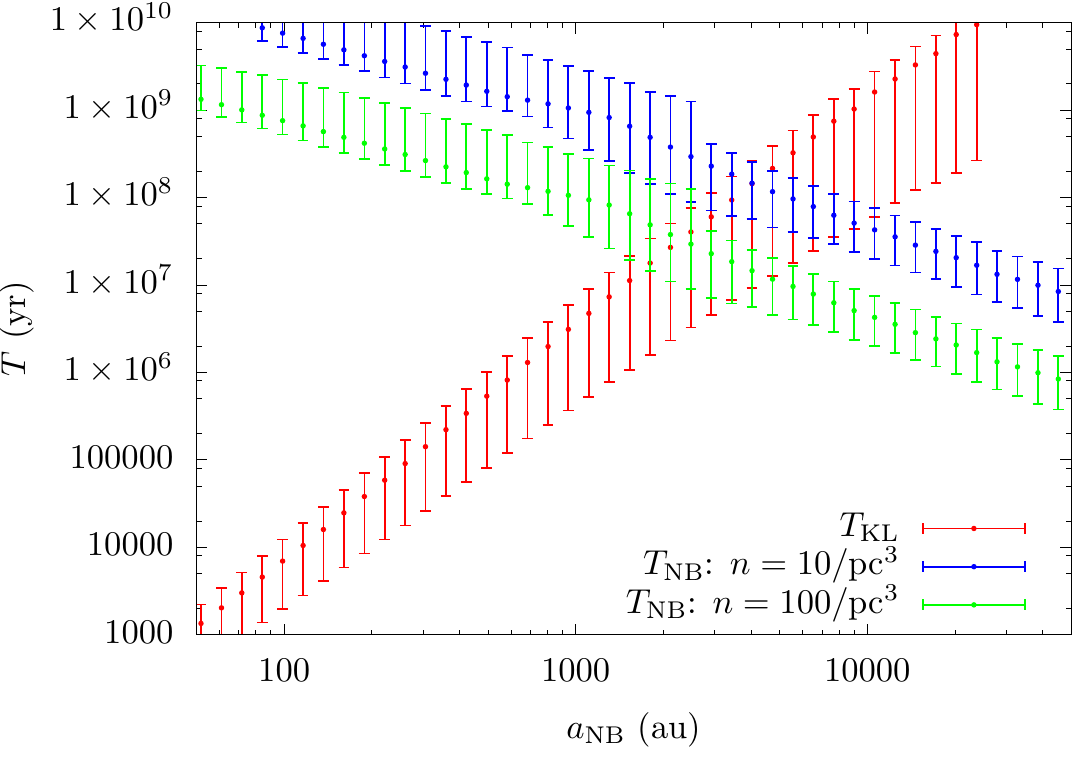}
\caption{Timescales for the SJ-companion system as a function of the separation between the SJ-pair and the companion. Red is for the Kozai--Lidov mechanism induced by the companion on Jupiter $T_\mathrm{KL}$. Blue and green are for the breakup of the system due to scattering another star $T_\mathrm{NB}$, for density $n=10$ or 100/pc$^3$, respectively.}
\label{fig-twotimescale}
\end{figure}

In Figure \ref{fig-twotimescale}, we plot, as a function of the separation between the SJ-pair and the companion star, in red the timescale of the Kozai--Lidov mechanism $T_\mathrm{KL}$ and in blue and green the timescale for the breakup of the SJ-companion system $T_\mathrm{NB}$ for stellar density $n=10$ and 100/pc$^3$; the error bars represent the 16th and 84th percentiles. As expected, $T_\mathrm{KL}$ is increasing steeply from $\sim$ 1000s of yr for $a_\mathrm{NB}\sim$ 100s of au to the age of the universe at $a_\mathrm{NB}>10^4$s of au. $T_\mathrm{NB}$ is decreasing more slowly and that for $n=$10/pc$^3$ larger than for 100/pc$^3$ by a factor of 10. Though the dispersion is large, overall, for $a_\mathrm{NB}$ lower than a few 1000s of au, $T_\mathrm{NB}>T_\mathrm{KL}$, meaning that Kozai--Lidov mechanism may be in effect. Figure \ref{fig-sjtejupijup} after debiasing, $a_\mathrm{NB}$ for the SJ-companion system centred around a few 1000s of au, coincident with where $T_\mathrm{NB}$ and $T_\mathrm{KL}$ are close. Therefore, very roughly, for half of the SJ-companion systems created via scatterings between the SJ-pair and a binary, the companion is able to excite Jupiter's orbit through the Kozai--Lidov mechanism before it is stripped by another scattering event.

For the Kozai--Lidov mechanism to operate, we require that $T_\mathrm{NB}>T_\mathrm{KL}$ for either of the two stellar densities and that the relative inclination between the planetary orbit and the companion orbit should be in the range $(40^\circ,140^\circ)$; we refer to these as KL-excitation. The resulting cross sections are listed in the bottom two rows of Table \ref{tab-cross}. The cross sections for $n=10$ or 100/pc$^3$ differ by 40\% and in both cases, that for KL-excitation is a few times that of the cross section for the group of five possible outcomes collectively labelled in Table \ref{tab-cross} as ``instability''. As before, an effective cross section has been calculated for a binary fraction of $f_\mathrm{bin}=0.5$. In general, KL-excitation is 50 per cent as likely as the formation of SJ-companion. Here we refer to the systems that are exposed to Kozai--Lidov mechanism as being damaged because they may be subject to the instability so-induced \citep[see for example][]{Malmberg2007a}. Then a re-examination of Figure \ref{fig-bin_vs_sin} implies, bearing in mind that the cross section for Kozai--Lidov damage is half of that for SJ-companion, Kozai--Lidov damage is more effective than immediate ejection during the scattering as long as the binary fraction is larger than a few times 0.01 (as the green line has a value more than double of the  solid black line in Figure \ref{fig-bin_vs_sin}).

The above inference should be treated with caution. Here, $T_\mathrm{NB}$ possibly represents an upper limit in that scatterings with binaries are not considered but those may destroy the SJ-companion system more effectively. On the other hand, we have only considered the disruption of the SJ-companion system where as an encounter may also either harden the system or increase the eccentricity, both reducing the Kozai--Lidov timescale. Moreover, the two timescales at $a_\mathrm{NB}\sim$ 1000s of au are of the order of $10^8$ yr, not hugely shorter than the lifetimes of small clusters themselves \citep{Adams2001a,Lamers2006}. So it is then possible that the cluster dissolves more quickly than the SJ-companion system experiences a disrupting encounter.

Finally, the fact that $T_\mathrm{NB}$ can be only a few Myr at $a_\mathrm{bin}=10^4$ au in clusters of stellar density $n=100$/pc$^3$ raises the concern whether our initial binary population is reasonable in that wide binaries could have been disrupted in a few crossing times before they scatter with the SJ-pair \citep{Kroupa1995,Parker2009}. In our Monte Carlo scattering simulations, wide binaries $a_\mathrm{bin}>1000$ au form the tail of the lognormal distribution of $P_\mathrm{bin}$, accounting for 13\% of the total binary population. We reexamine our simulations with $f_\mathrm{bin}=1$, now removing contribution from those of $a_\mathrm{bin}>1000$ au, and calculate the corresponding cross sections. Compared to the original binary population ($a_\mathrm{bin}<10000$ au), the cross section for instability decreases by slightly 10\%. Then that for the formation of the SJ-companion configuration drops by 50\%, because the wide binaries $a_\mathrm{bin}>1000$ au that can give rise to this outcome at large impact parameters are removed.

\section{Implication for the solar system}\label{sec-solar}
The solar system itself is believed to have originated from a cluster \citep[see][and reference therein]{Adams2010}. For the solar system to be contaminated by radioactive elements, it has to reside close to a massive star but such stars are rare and only populous clusters contain them; this could be used to put a lower limit on the cluster size \citep{Adams2001,Parker2014}. On the other hand, larger cluster sizes mean higher background UV that may photo-evaporate the protoplanetary disk, hindering planet formation \citep{Adams2006,Winter2018} and higher frequency of close encounters which may destroy the already-formed planetary systems \citep{Adams2001,Li2015}; these two processes can be used to put an upper limit on the cluster size \citep{Adams2006}. Working from both ends, the solar system birth cluster was estimated to host a few 1000s of member stars \citep{Adams2010}. Here with the updated cross sections, we would like to test if more stringent constraints can be put on that cluster.

We calculate the cross section $\sigma$ for the moderate disruption of the SJ-pair, including Jupiter's instability (Table \ref{tab-cross}) plus its eccentricity excited to values larger than 0.1 (whatever the outcome), as done in \citet{Li2015}. From our Monte Carlo simulations, $\langle\sigma\rangle_\mathrm{eff}$ is $2.25\times10^5$ au$^2$ for $f_\mathrm{bin}=0.5$, twice of that for Jupiter's direct ejection. We stress that this is a lower limit as the way our simulations is designed only guarantees the convergence for the cross section for SJ-change (instability+SJ-companion) but not for the planet's eccentricity excitation. Also, later pumping through Kozai--Lidov mechanism is not included here.

Then if this $\langle\sigma\rangle_\mathrm{eff}$ also scales linearly with the planetary semimajor axis \citep{Li2015}, arguably Neptune would be 6 times as vulnerable as Jupiter, leading to a cross section of $1.35\times10^6$ au$^2$. With Equation \eqref{eq-rate}, we can derive the relative occurrence rate. And combined with the actual rates for planet ejection in cluster simulations \citep[e.g.,][]{Fujii2019} (for details, cf. Section \ref{sec-sec-occ}), we estimate that in our fiducial cluster of stellar density of $n=50$ pc$^{-3}$ and lifetime $T_\mathrm{c}=100$ Myr, a planet at 30 au would be unstable from the host or excited to an eccentricity higher than 0.1 at a chance of 15\%. So the Neptunian observed cold orbit is not inconsistent with Solar system originating from an open cluster. Furthermore, if the encounter is early, a massive Kuiper belt can easily damp an eccentricity of 0.1.

 Those Kuiper belt objects themselves, especially those on cold orbits beyond $\sim40$ au, would be inevitably perturbed by the scattering. The same reasoning as used above for Neptune indicates a cross section of $1.8\times10^6$ au$^2$ for an eccentricity excitation higher than 0.1 and an associated chance of 20\% in the afore-discussed cluster. This puts a slightly more stringent constraint on the birth cluster of the solar system. However, we also note that these Kuiper belt objects are subject to later evolution and those excited may be removed due to resonance crossing \citep{Morbidelli2014}.

 Among all the Kuiper belt objects, Sedna-like objects stand out as being also potential members of the inner Oort cloud. These objects have very wide, eccentric orbits generally beyond the reach of Neptune. Their formation has been, among the other models, often attributed to a stellar flyby \citep{Morbidelli2004}. Two scenarios are possible: (1) they are initially revolving around the Sun on circular orbits and are flung onto elongated orbits by the intruding star; (2) they are captured by the Sun from the intruder. Model (1) has been discussed thoroughly as being perturbed by a single star \citep[e.g.,][]{Pfalzner2018}. Here in this work, we have shown binaries, compared to single stars, greatly enhance the chance of ejection of solar system objects. Hence, it the eccentricity excitation follows a similar story \citep{Li2015}, binaries would significantly increase the chance of creating such highly excited objects. As for model (2), if binaries are involved, it could be that the Sun encounters the binary or the Sun is itself a component of the binary. It seems that in both two cases, the capture cross section for the Sun to capture may have been boosted by a factor of at most a few compared to single-single encounters as from our work and from \citet[][but see also \citealt{Siraj2020}]{Wang2020}.

Exchange of the solar system a whole acquires a stellar companion is 10 times more likely as to eject Jupiter. The fact that such a companion tends to be 1000s of au from the Sun (top panel of Figure \ref{fig-sjtejupijup}) suggests it is relatively easy to break up and its interaction between the solar system objects is probably secular. As a consequence, the solar system could hold together and be tilted through Kozai--Lidov mechanism \citep[e.g.,][but see also \citealt{Malmberg2007a}]{Innanen1997}, possibly against the equator of the Sun \citep{Gomes2016}. A detailed account on the implications of a stellar companion of the solar system the is beyond the scope of this work. We emphasise that this scenario is not inconsistent and may be helpful in explaining the outer solar system features and we refer to \citet{Siraj2020} for a recent extended discussion.

\section{Conclusions}\label{sec-con}
Stars are often born in a clustered environment with many stellar siblings. Clusters of 100s of stars or more can hold together for long enough and the members will interact with each other through close encounters. Then inevitably, the planets orbiting those stars are subject to flyby encounters. Such encounters not only involve single stars but also binaries. In this work, we study the scatterings between a planetary system (exemplified by the Sun-Jupiter pair) and a binary.

In our first set of simulations, we have picked binary parameters from a grid, varying their total mass, the mass ratio, the semimajor axis, the eccentricity and the encounter velocity. We derive the largest impact parameters where the configuration of the Sun-Jupiter pair changes and find that it can be fitted as a power-law function of the binary parameters. Then the cross section for the ejection of Jupiter has been estimated for each binary parameter, showing that the more massive the binary mass, the smaller the encounter velocity and the wider the binary separation, the larger the cross section. 

In our second set, we create realistic binary parameters as drawn from the observed population. Here, large-enough impact parameters as derived from the previous grid-simulations are used, assuring the convergence for our so-obtained cross sections. Additionally, another set of simulations for encountering singles are done. The two enables us to derive an effective cross section for a stellar population of both single and binary stars.

Scatterings between binaries and a planetary system encompass rich dynamics and the outcomes are diverse, including the planet's ejection, collision, capture, etc. In general, binaries lead to cross sections for Jupiter's ejection and collision larger than those by single stars by an order of magnitude or more. Hence, as long as the binary fraction of a cluster is larger than $\sim0.1$, it is the binaries that dominate both the planet's ejection and collision. More frequently by an order of magnitude, the Sun-Jupiter pair remains intact and acquires a stellar companion. Such a companion is most likely 1000s of au from the Sun and for half of the so-formed systems, the companion can excite Jupiter's orbit through Kozai--Lidov mechanism before it is stripped in later scatterings. This means that this long-term Kozai--Lidov excitation of the SJ-pair has a cross section several times that of immediate ejection during the scattering. For a fiducial cluster of a stellar density $50$ pc$^{-3}$, a lifetime of 100 Myr and a binary fraction of 0.5, we estimate that of the order of 0.01 of the Jupiters are ejected, 0.001 collide with a star, 0.001 are captured by other stars and 0.1 of the Sun-Jupiter pairs have picked up a companion.

Finally, we discuss the implications for the solar system in its birth cluster. We show that Neptune might be ejected or excited to an eccentricity $>0.1$ due to the encounter flybys at a probability of 0.15. Also, we show that the solar system may once have a stellar companion via scattering a binary. That companion is typically, as described above, 1000s of au wide and hence, the interaction between it and the solar system objects is secular and it can be stripped in later stellar encounters.

\section*{Acknowledgements}

The authors thank an anonymous referee for helpful comments. D.L. acknowledges financial support from Knut and Alice Wallenberg Foundation (2014.0017 and 2012.0150), from Vetenskapsrådet (2017-04945), and from the Royal Physiographic Society of Lund (F 2019/769). Computations were carried out at the center for scientific and technical computing at Lund University (LUNARC) through the Swedish National Infrastructure for Computing (SNIC) via project 2019/3-398.

\section*{Data Availability}
The data underlying this article will be shared on reasonable request to the corresponding author.



\begin{thebibliography}{}
\makeatletter
\relax
\def\mn@urlcharsother{\let\do\@makeother \do\$\do\&\do\#\do\^\do\_\do\%\do\~}
\def\mn@doi{\begingroup\mn@urlcharsother \@ifnextchar [ {\mn@doi@}
  {\mn@doi@[]}}
\def\mn@doi@[#1]#2{\def\@tempa{#1}\ifx\@tempa\@empty \href
  {http://dx.doi.org/#2} {doi:#2}\else \href {http://dx.doi.org/#2} {#1}\fi
  \endgroup}
\def\mn@eprint#1#2{\mn@eprint@#1:#2::\@nil}
\def\mn@eprint@arXiv#1{\href {http://arxiv.org/abs/#1} {{\tt arXiv:#1}}}
\def\mn@eprint@dblp#1{\href {http://dblp.uni-trier.de/rec/bibtex/#1.xml}
  {dblp:#1}}
\def\mn@eprint@#1:#2:#3:#4\@nil{\def\@tempa {#1}\def\@tempb {#2}\def\@tempc
  {#3}\ifx \@tempc \@empty \let \@tempc \@tempb \let \@tempb \@tempa \fi \ifx
  \@tempb \@empty \def\@tempb {arXiv}\fi \@ifundefined
  {mn@eprint@\@tempb}{\@tempb:\@tempc}{\expandafter \expandafter \csname
  mn@eprint@\@tempb\endcsname \expandafter{\@tempc}}}

\bibitem[\protect\citeauthoryear{Adams}{Adams}{2010}]{Adams2010}
Adams F.~C.,  2010, \mn@doi [Annual Review of Astronomy and Astrophysics]
  {10.1146/annurev-astro-081309-130830}, 48, 47

\bibitem[\protect\citeauthoryear{Adams \& Laughlin}{Adams \&
  Laughlin}{2001}]{Adams2001}
Adams F.~C.,  Laughlin G.,  2001, \mn@doi [Icarus] {10.1006/icar.2000.6567},
  150, 151

\bibitem[\protect\citeauthoryear{Adams \& Myers}{Adams \&
  Myers}{2001}]{Adams2001a}
Adams F.~C.,  Myers P.~C.,  2001, \mn@doi [The Astrophysical Journal]
  {10.1086/320941}, 553, 744

\bibitem[\protect\citeauthoryear{Adams, Proszkow, Fatuzzo  \& Myers}{Adams
  et~al.}{2006}]{Adams2006}
Adams F.~C.,  Proszkow E.~M.,  Fatuzzo M.,   Myers P.~C.,  2006, \mn@doi [The
  Astrophysical Journal] {10.1086/500393}, 641, 504

\bibitem[\protect\citeauthoryear{Antognini \& Thompson}{Antognini \&
  Thompson}{2016}]{Antognini2016}
Antognini J. M.~O.,  Thompson T.~A.,  2016, \mn@doi [Monthly Notices of the
  Royal Astronomical Society] {10.1093/mnras/stv2938}, 456, 4219

\bibitem[\protect\citeauthoryear{Bacon, Sigurdsson  \& Davies}{Bacon
  et~al.}{1996}]{Bacon1996}
Bacon D.,  Sigurdsson S.,   Davies M.~B.,  1996, \mn@doi [Monthly Notices of
  the Royal Astronomical Society] {10.1093/mnras/281.3.830}, 281, 830

\bibitem[\protect\citeauthoryear{Bailey et~al.,}{Bailey
  et~al.}{2014}]{Bailey2014}
Bailey V.,  et~al., 2014, \mn@doi [Astrophysical Journal Letters]
  {10.1088/2041-8205/780/1/L4}, 780

\bibitem[\protect\citeauthoryear{Battinelli \& Capuzzo-Dolcetta}{Battinelli \&
  Capuzzo-Dolcetta}{1991}]{Battinelli1991}
Battinelli P.,  Capuzzo-Dolcetta R.,  1991, \mn@doi [Monthly Notices of the
  Royal Astronomical Society] {10.1093/mnras/249.1.76}, 249, 76

\bibitem[\protect\citeauthoryear{Bica \& Bonatto}{Bica \&
  Bonatto}{2005}]{Bica2005}
Bica E.,  Bonatto C.,  2005, \mn@doi [Astronomy {\&} Astrophysics]
  {10.1051/0004-6361:20042023}, 431, 943

\bibitem[\protect\citeauthoryear{Binney \& Tremaine}{Binney \&
  Tremaine}{2008}]{Binney2008}
Binney J.,  Tremaine S.,  2008, {Galactic dynamics}.
Princeton University Press, \url
  {http://adsabs.harvard.edu/abs/2008gady.book.....B}

\bibitem[\protect\citeauthoryear{Bouvier, Rigaut  \& Nadeau}{Bouvier
  et~al.}{1997}]{Bouvier1997}
Bouvier J.,  Rigaut F.,   Nadeau D.,  1997, Astronomy and Astrophysics, 323,
  139

\bibitem[\protect\citeauthoryear{Cai, Kouwenhoven, Zwart  \& Spurzem}{Cai
  et~al.}{2017}]{Cai2017}
Cai M.~X.,  Kouwenhoven M. B.~N.,  Zwart S. F.~P.,   Spurzem R.,  2017, \mn@doi
  [Monthly Notices of the Royal Astronomical Society] {10.1093/mnras/stx1464},
  470, 4337

\bibitem[\protect\citeauthoryear{Duquennoy \& Mayor}{Duquennoy \&
  Mayor}{1991}]{Duquennoy1991}
Duquennoy A.,  Mayor M.,  1991, Astronomy and astrophysics, 248, 485

\bibitem[\protect\citeauthoryear{Eggenberger \& Udry}{Eggenberger \&
  Udry}{2010}]{Eggenberger2010}
Eggenberger A.,  Udry S.,  2010, pp 19--49,
  \mn@doi{10.1007/978-90-481-8687-7_2}, \url
  {http://link.springer.com/10.1007/978-90-481-8687-7{\_}2}

\bibitem[\protect\citeauthoryear{Elmegreen \& Clemens}{Elmegreen \&
  Clemens}{1985}]{Elmegreen1985}
Elmegreen B.~G.,  Clemens C.,  1985, \mn@doi [The Astrophysical Journal]
  {10.1086/163320}, 294, 523

\bibitem[\protect\citeauthoryear{Fabrycky \& Tremaine}{Fabrycky \&
  Tremaine}{2007}]{Fabrycky2007}
Fabrycky D.,  Tremaine S.,  2007, \mn@doi [The Astrophysical Journal]
  {10.1086/521702}, 669, 1298

\bibitem[\protect\citeauthoryear{Farago \& Laskar}{Farago \&
  Laskar}{2010}]{Farago2010}
Farago F.,  Laskar J.,  2010, \mn@doi [Monthly Notices of the Royal
  Astronomical Society] {10.1111/j.1365-2966.2009.15711.x}, 401, 1189

\bibitem[\protect\citeauthoryear{Fragione}{Fragione}{2019}]{Fragione2019}
Fragione G.,  2019, \mn@doi [Monthly Notices of the Royal Astronomical Society]
  {10.1093/mnras/sty3367}, 483, 3465

\bibitem[\protect\citeauthoryear{Fregeau, Cheung, {Portegies Zwart}  \&
  Rasio}{Fregeau et~al.}{2004}]{Fregeau2004}
Fregeau J.~M.,  Cheung P.,  {Portegies Zwart} S.~F.,   Rasio F.~A.,  2004,
  \mn@doi [Monthly Notices of the Royal Astronomical Society]
  {10.1111/j.1365-2966.2004.07914.x}, 352, 1

\bibitem[\protect\citeauthoryear{Fregeau, Chatterjee  \& Rasio}{Fregeau
  et~al.}{2006}]{Fregeau2006}
Fregeau J.~M.,  Chatterjee S.,   Rasio F.~A.,  2006, \mn@doi [The Astrophysical
  Journal] {10.1086/500111}, 640, 1086

\bibitem[\protect\citeauthoryear{Fujii \& Hori}{Fujii \&
  Hori}{2019}]{Fujii2019}
Fujii M.,  Hori Y.,  2019, \mn@doi [Astronomy {\&} Astrophysics]
  {10.1051/0004-6361/201834677}, 624, A110

\bibitem[\protect\citeauthoryear{Geller \& Leigh}{Geller \&
  Leigh}{2015}]{Geller2015}
Geller A.~M.,  Leigh N. W.~C.,  2015, \mn@doi [The Astrophysical Journal
  Letters] {10.1088/2041-8205/808/1/L25}, 808, L25

\bibitem[\protect\citeauthoryear{Gomes, Deienno  \& Morbidelli}{Gomes
  et~al.}{2016}]{Gomes2016}
Gomes R.,  Deienno R.,   Morbidelli A.,  2016, \mn@doi [The Astronomical
  Journal] {10.3847/1538-3881/153/1/27}, 153, 27

\bibitem[\protect\citeauthoryear{Goodwin, Kroupa, Goodman  \& Burkert}{Goodwin
  et~al.}{2007}]{Goodwin2007}
Goodwin S.~P.,  Kroupa P.,  Goodman A.,   Burkert A.,  2007, in Reipurth V B.,
  Jewitt D.,   Keil K.,  eds, , Protostars and Planets V.
University of Arizona Press, Tucson, pp 133--147 (\mn@eprint {arXiv}
  {0603233}), \url {http://arxiv.org/abs/astro-ph/0603233}

\bibitem[\protect\citeauthoryear{Hao, Kouwenhoven  \& Spurzem}{Hao
  et~al.}{2013}]{Hao2013}
Hao W.,  Kouwenhoven M.~B.,   Spurzem R.,  2013, \mn@doi [Monthly Notices of
  the Royal Astronomical Society] {10.1093/mnras/stt771}, 433, 867

\bibitem[\protect\citeauthoryear{Heggie, Hut  \& McMillan}{Heggie
  et~al.}{1996}]{Heggie1996}
Heggie D.~C.,  Hut P.,   McMillan S. L.~W.,  1996, \mn@doi [The Astrophysical
  Journal] {10.1086/177611}, 467, 359

\bibitem[\protect\citeauthoryear{Hills \& Dissly}{Hills \&
  Dissly}{1989}]{Hills1989}
Hills J.~G.,  Dissly R.~W.,  1989, \mn@doi [The Astronomical Journal]
  {10.1086/115197}, 98, 1069

\bibitem[\protect\citeauthoryear{Holman \& Wiegert}{Holman \&
  Wiegert}{1999}]{Holman1999}
Holman M.~J.,  Wiegert P.~A.,  1999, \mn@doi [The Astronomical Journal]
  {10.1086/300695}, 117, 621

\bibitem[\protect\citeauthoryear{Hut}{Hut}{1983}]{Hut1983a}
Hut P.,  1983, \mn@doi [The Astrophysical Journal] {10.1086/160957}, 268, 342

\bibitem[\protect\citeauthoryear{Hut \& Bahcall}{Hut \&
  Bahcall}{1983}]{Hut1983}
Hut P.,  Bahcall J.~N.,  1983, \mn@doi [The Astrophysical Journal]
  {10.1086/160956}, 268, 319

\bibitem[\protect\citeauthoryear{Innanen, Zheng, Mikkola  \& Valtonen}{Innanen
  et~al.}{1997}]{Innanen1997}
Innanen K.~a.,  Zheng J.~Q.,  Mikkola S.,   Valtonen M.~J.,  1997, \mn@doi [The
  Astronomical Journal] {10.1086/118405}, 113, 1915

\bibitem[\protect\citeauthoryear{Jaehnig, {Da Rio}  \& Tan}{Jaehnig
  et~al.}{2015}]{Jaehnig2015}
Jaehnig K.~O.,  {Da Rio} N.,   Tan J.~C.,  2015, \mn@doi [The Astrophysical
  Journal] {10.1088/0004-637X/798/2/126}, 798, 126

\bibitem[\protect\citeauthoryear{J{\'{i}}lkov{\'{a}}, Hamers, Hammer  \&
  Zwart}{J{\'{i}}lkov{\'{a}} et~al.}{2016}]{Jilkova2016}
J{\'{i}}lkov{\'{a}} L.,  Hamers A.~S.,  Hammer M.,   Zwart S.~P.,  2016,
  \mn@doi [Monthly Notices of the Royal Astronomical Society]
  {10.1093/mnras/stw264}, 457, 4218

\bibitem[\protect\citeauthoryear{Kiseleva, Eggleton  \& Mikkola}{Kiseleva
  et~al.}{1998}]{Kiseleva1998}
Kiseleva L.~G.,  Eggleton P.~P.,   Mikkola S.,  1998, \mn@doi [Monthly Notices
  of the Royal Astronomical Society] {10.1046/j.1365-8711.1998.01903.x}, 300,
  292

\bibitem[\protect\citeauthoryear{Kouwenhoven, Goodwin, Parker, Davies, Malmberg
   \& Kroupa}{Kouwenhoven et~al.}{2010}]{Kouwenhoven2010}
Kouwenhoven M. B.~N.,  Goodwin S.~P.,  Parker R.~J.,  Davies M.~B.,  Malmberg
  D.,   Kroupa P.,  2010, \mn@doi [Monthly Notices of the Royal Astronomical
  Society] {10.1111/j.1365-2966.2010.16399.x}

\bibitem[\protect\citeauthoryear{Kozai}{Kozai}{1962}]{Kozai1962}
Kozai Y.,  1962, \mn@doi [The Astronomical Journal] {10.1086/108876}, 67, 579

\bibitem[\protect\citeauthoryear{Kraus, Ireland, Cieza, Hinkley, Dupuy, Bowler
  \& Liu}{Kraus et~al.}{2013}]{Kraus2013}
Kraus A.~L.,  Ireland M.~J.,  Cieza L.~A.,  Hinkley S.,  Dupuy T.~J.,  Bowler
  B.~P.,   Liu M.~C.,  2013, \mn@doi [The Astrophysical Journal]
  {10.1088/0004-637X/781/1/20}, 781, 20

\bibitem[\protect\citeauthoryear{Kraus, Ireland, Huber, Mann  \& Dupuy}{Kraus
  et~al.}{2016}]{Kraus2016}
Kraus A.~L.,  Ireland M.~J.,  Huber D.,  Mann A.~W.,   Dupuy T.~J.,  2016,
  \mn@doi [The Astronomical Journal] {10.3847/0004-6256/152/1/8}, 152, 8

\bibitem[\protect\citeauthoryear{Kroupa}{Kroupa}{1995}]{Kroupa1995}
Kroupa P.,  1995, \mn@doi [Monthly Notices of the Royal Astronomical Society]
  {10.1093/mnras/277.4.1491}, 277, 1491

\bibitem[\protect\citeauthoryear{Kroupa}{Kroupa}{2001}]{Kroupa2001}
Kroupa P.,  2001, \mn@doi [Monthly Notices of the Royal Astronomical Society]
  {10.1046/j.1365-8711.2001.04022.x}, 322, 231

\bibitem[\protect\citeauthoryear{Kroupa \& Burkert}{Kroupa \&
  Burkert}{2001}]{Kroupa2001a}
Kroupa P.,  Burkert A.,  2001, \mn@doi [The Astrophysical Journal]
  {10.1086/321515}, 555, 945

\bibitem[\protect\citeauthoryear{Lada \& Lada}{Lada \& Lada}{2003}]{Lada2003}
Lada C.~J.,  Lada E.~A.,  2003, \mn@doi [Annual Review of Astronomy and
  Astrophysics] {10.1146/annurev.astro.41.011802.094844}, 41, 57

\bibitem[\protect\citeauthoryear{Lamers \& Gieles}{Lamers \&
  Gieles}{2006}]{Lamers2006}
Lamers H. J. G. L.~M.,  Gieles M.,  2006, \mn@doi [Astronomy {\&} Astrophysics]
  {10.1051/0004-6361:20065567}, 455, L17

\bibitem[\protect\citeauthoryear{Laughlin \& Adams}{Laughlin \&
  Adams}{1998}]{Laughlin1998}
Laughlin G.,  Adams F.,  1998, \mn@doi [The Astrophysical Journal]
  {10.1086/311736}, 508, L171

\bibitem[\protect\citeauthoryear{Li \& Adams}{Li \& Adams}{2015}]{Li2015}
Li G.,  Adams F.~C.,  2015, \mn@doi [Monthly Notices of the Royal Astronomical
  Society] {10.1093/mnras/stv012}, 448, 344

\bibitem[\protect\citeauthoryear{Li, Mustill  \& Davies}{Li
  et~al.}{2019}]{Li2019}
Li D.,  Mustill A.~J.,   Davies M.~B.,  2019, \mn@doi [Monthly Notices of the
  Royal Astronomical Society] {10.1093/mnras/stz1794}, 488, 1366

\bibitem[\protect\citeauthoryear{Li, Mustill  \& Davies}{Li
  et~al.}{2020}]{Li2020a}
Li D.,  Mustill A.~J.,   Davies M.~B.,  2020, \mn@doi [Monthly Notices of the
  Royal Astronomical Society] {10.1093/mnras/staa1622}, 496, 1149

\bibitem[\protect\citeauthoryear{Lidov}{Lidov}{1962}]{Lidov1962}
Lidov M.,  1962, \mn@doi [Planetary and Space Science]
  {10.1016/0032-0633(62)90129-0}, 9, 719

\bibitem[\protect\citeauthoryear{Malmberg, Davies  \& Chambers}{Malmberg
  et~al.}{2007a}]{Malmberg2007a}
Malmberg D.,  Davies M.~B.,   Chambers J.~E.,  2007a, \mn@doi [Monthly Notices
  of the Royal Astronomical Society: Letters]
  {10.1111/j.1745-3933.2007.00291.x}, 377, L1

\bibitem[\protect\citeauthoryear{Malmberg, {De Angeli}, Davies, Church, MacKey
  \& Wilkinson}{Malmberg et~al.}{2007b}]{Malmberg2007}
Malmberg D.,  {De Angeli} F.,  Davies M.~B.,  Church R.~P.,  MacKey D.,
  Wilkinson M.~I.,  2007b, \mn@doi [Monthly Notices of the Royal Astronomical
  Society] {10.1111/j.1365-2966.2007.11885.x}, 378, 1207

\bibitem[\protect\citeauthoryear{Malmberg, Davies  \& Heggie}{Malmberg
  et~al.}{2011}]{Malmberg2011}
Malmberg D.,  Davies M.~B.,   Heggie D.~C.,  2011, \mn@doi [Monthly Notices of
  the Royal Astronomical Society] {10.1111/j.1365-2966.2010.17730.x}, 411, 859

\bibitem[\protect\citeauthoryear{Marks \& Kroupa}{Marks \&
  Kroupa}{2012}]{Marks2012}
Marks M.,  Kroupa P.,  2012, \mn@doi [Astronomy {\&} Astrophysics]
  {10.1051/0004-6361/201118231}, 543, A8

\bibitem[\protect\citeauthoryear{Martin}{Martin}{2018}]{Martin2018}
Martin D.~V.,  2018, in , Handbook of Exoplanets.
Springer International Publishing, Cham, pp 1--26,
  \mn@doi{10.1007/978-3-319-30648-3_156-1}, \url
  {http://link.springer.com/10.1007/978-3-319-30648-3{\_}156-1}

\bibitem[\protect\citeauthoryear{Milone, Piotto, Bedin  \& Sarajedini}{Milone
  et~al.}{2008}]{Milone2008}
Milone A.~P.,  Piotto G.,  Bedin L.~R.,   Sarajedini A.,  2008, in Cassisi S.,
  Salaris M.,  eds,  Vol. 79, Memorie della Societ{\`{a}} Astronomica Italiana.
  p.~623 (\mn@eprint {arXiv} {0801.3177}), \url
  {http://arxiv.org/abs/0801.3177}

\bibitem[\protect\citeauthoryear{Morbidelli \& Levison}{Morbidelli \&
  Levison}{2004}]{Morbidelli2004}
Morbidelli A.,  Levison H.~F.,  2004, \mn@doi [The Astronomical Journal]
  {10.1086/424617}, 128, 2564

\bibitem[\protect\citeauthoryear{Morbidelli, Gaspar  \& Nesvorny}{Morbidelli
  et~al.}{2014}]{Morbidelli2014}
Morbidelli A.,  Gaspar H.~S.,   Nesvorny D.,  2014, \mn@doi [Icarus]
  {10.1016/j.icarus.2013.12.023}, 232, 81

\bibitem[\protect\citeauthoryear{Mustill, Raymond  \& Davies}{Mustill
  et~al.}{2016}]{Mustill2016}
Mustill A.~J.,  Raymond S.~N.,   Davies M.~B.,  2016, \mn@doi [Monthly Notices
  of the Royal Astronomical Society: Letters] {10.1093/mnrasl/slw075}, 460,
  L109

\bibitem[\protect\citeauthoryear{Parker, Goodwin, Kroupa  \&
  Kouwenhoven}{Parker et~al.}{2009}]{Parker2009}
Parker R.~J.,  Goodwin S.~P.,  Kroupa P.,   Kouwenhoven M. B.~N.,  2009,
  \mn@doi [Monthly Notices of the Royal Astronomical Society]
  {10.1111/j.1365-2966.2009.15032.x}, 397, 1577

\bibitem[\protect\citeauthoryear{Parker, Church, Davies  \& Meyer}{Parker
  et~al.}{2014}]{Parker2014}
Parker R.~J.,  Church R.~P.,  Davies M.~B.,   Meyer M.~R.,  2014, \mn@doi
  [Monthly Notices of the Royal Astronomical Society] {10.1093/mnras/stt1957},
  437, 946

\bibitem[\protect\citeauthoryear{Pfalzner, Vogel, Scharw{\"{a}}chter  \&
  Olczak}{Pfalzner et~al.}{2005}]{Pfalzner2005}
Pfalzner S.,  Vogel P.,  Scharw{\"{a}}chter J.,   Olczak C.,  2005, \mn@doi
  [Astronomy {\&} Astrophysics] {10.1051/0004-6361:20042467}, 437, 967

\bibitem[\protect\citeauthoryear{Pfalzner, Bhandare, Vincke  \&
  Lacerda}{Pfalzner et~al.}{2018}]{Pfalzner2018}
Pfalzner S.,  Bhandare A.,  Vincke K.,   Lacerda P.,  2018, \mn@doi [The
  Astrophysical Journal] {10.3847/1538-4357/aad23c}, 863, 45

\bibitem[\protect\citeauthoryear{Proszkow \& Adams}{Proszkow \&
  Adams}{2009}]{Proszkow2009}
Proszkow E.-M.,  Adams F.~C.,  2009, \mn@doi [The Astrophysical Journal
  Supplement Series] {10.1088/0067-0049/185/2/486}, 185, 486

\bibitem[\protect\citeauthoryear{Raghavan et~al.,}{Raghavan
  et~al.}{2010}]{Raghavan2010}
Raghavan D.,  et~al., 2010, \mn@doi [Astrophysical Journal, Supplement Series]
  {10.1088/0067-0049/190/1/1}, 190, 1

\bibitem[\protect\citeauthoryear{Richichi, Chen, Cusano, Fors, Moerchen  \&
  Komonjinda}{Richichi et~al.}{2012}]{Richichi2012}
Richichi A.,  Chen W.~P.,  Cusano F.,  Fors O.,  Moerchen M.,   Komonjinda S.,
  2012, \mn@doi [Astronomy {\&} Astrophysics] {10.1051/0004-6361/201219041},
  541, A96

\bibitem[\protect\citeauthoryear{Rodet, Beust, Bonnefoy, Lagrange, Galli,
  Ducourant  \& Teixeira}{Rodet et~al.}{2017}]{Rodet2017}
Rodet L.,  Beust H.,  Bonnefoy M.,  Lagrange A.~M.,  Galli P. A.~B.,  Ducourant
  C.,   Teixeira R.,  2017, \mn@doi [Astronomy {\&} Astrophysics]
  {10.1051/0004-6361/201630269}, 602, A12

\bibitem[\protect\citeauthoryear{Schwarz, Funk, Zechner  \&
  Bazs{\'{o}}}{Schwarz et~al.}{2016}]{Schwarz2016a}
Schwarz R.,  Funk B.,  Zechner R.,   Bazs{\'{o}} 2016, \mn@doi [Monthly Notices
  of the Royal Astronomical Society] {10.1093/mnras/stw1218}, 460, 3598

\bibitem[\protect\citeauthoryear{Siraj \& Loeb}{Siraj \&
  Loeb}{2020}]{Siraj2020}
Siraj A.,  Loeb A.,  2020, \mn@doi [The Astrophysical Journal]
  {10.3847/2041-8213/abac66}, 899, L24

\bibitem[\protect\citeauthoryear{Sollima, Beccari, Ferraro, {Fusi Pecci}  \&
  Sarajedini}{Sollima et~al.}{2007}]{Sollima2007}
Sollima A.,  Beccari G.,  Ferraro F.~R.,  {Fusi Pecci} F.,   Sarajedini A.,
  2007, \mn@doi [Monthly Notices of the Royal Astronomical Society]
  {10.1111/j.1365-2966.2007.12116.x}, 380, 781

\bibitem[\protect\citeauthoryear{Sollima, Carballo-Bello, Beccari, Ferraro,
  Pecci  \& Lanzoni}{Sollima et~al.}{2010}]{Sollima2010}
Sollima A.,  Carballo-Bello J.~A.,  Beccari G.,  Ferraro F.~R.,  Pecci F.~F.,
  Lanzoni B.,  2010, \mn@doi [Monthly Notices of the Royal Astronomical
  Society] {10.1111/j.1365-2966.2009.15676.x}, 401, 577

\bibitem[\protect\citeauthoryear{Wang, Perna  \& Leigh}{Wang
  et~al.}{2020}]{Wang2020}
Wang Y.-H.,  Perna R.,   Leigh N. W.~C.,  2020, \mn@doi [Monthly Notices of the
  Royal Astronomical Society] {10.1093/mnras/staa1627}

\bibitem[\protect\citeauthoryear{Ward, Kruijssen  \& Rix}{Ward
  et~al.}{2020}]{Ward2020}
Ward J.~L.,  Kruijssen J. M.~D.,   Rix H.-W.,  2020, \mn@doi [Monthly Notices
  of the Royal Astronomical Society] {10.1093/mnras/staa1056}, 495, 663

\bibitem[\protect\citeauthoryear{White \& Ghez}{White \&
  Ghez}{2001}]{White2001}
White R.~J.,  Ghez A.~M.,  2001, \mn@doi [The Astrophysical Journal]
  {10.1086/321542}, 556, 265

\bibitem[\protect\citeauthoryear{Winter, Clarke, Rosotti, Ih, Facchini  \&
  Haworth}{Winter et~al.}{2018}]{Winter2018}
Winter A.~J.,  Clarke C.~J.,  Rosotti G.,  Ih J.,  Facchini S.,   Haworth
  T.~J.,  2018, \mn@doi [Monthly Notices of the Royal Astronomical Society]
  {10.1093/mnras/sty984}, 478, 2700

\bibitem[\protect\citeauthoryear{van~den Bergh}{van~den
  Bergh}{1981}]{VandenBergh1981}
van~den Bergh S.,  1981, \mn@doi [Publications of the Astronomical Society of
  the Pacific] {10.1086/130912}, 93, 712

\makeatother
\end{thebibliography}


\bsp    
\label{lastpage}
\end{document}